\documentclass[twocolumn,trackchanges]{aastex7}

\shorttitle{BEACON DR2: Galaxy Properties}
\shortauthors{Y. Zhang et al.}
\submitjournal{ApJ}

\begin{document}

\title{BEACON: JWST NIRCam Pure-parallel Imaging Survey. II. Physical Properties of $z=7-14$ Galaxies}

\correspondingauthor{Yechi Zhang}
\email{yechi@ipac.caltech.edu}
\author[0000-0003-3817-8739]{Yechi Zhang}
\affiliation{IPAC, California Institute of Technology, MC 314-6, 1200 E. California Boulevard, Pasadena, CA 91125, USA}
\email{yechi@ipac.caltech.edu}

\author[0000-0002-8512-1404]{Takahiro Morishita} 
\affiliation{IPAC, California Institute of Technology, MC 314-6, 1200 E. California Boulevard, Pasadena, CA 91125, USA}
\email{takahiro@ipac.caltech.edu}

\author[0009-0005-9953-433X]{Kimi C. Kreilgaard}
\affiliation{Cosmic Dawn Center (DAWN), Denmark}
\affiliation{Niels Bohr Institute, University of Copenhagen, Jagtvej 128, DK-2200 Copenhagen N, Denmark}
\email{kimi.cardoso.kreilgaard@nbi.ku.dk}

\author[0000-0002-3407-1785]{Charlotte A. Mason}
\affiliation{Cosmic Dawn Center (DAWN), Denmark}
\affiliation{Niels Bohr Institute, University of Copenhagen, Jagtvej 128, DK-2200 Copenhagen N, Denmark}
\email{charlotte.mason@nbi.ku.dk}

\author[0000-0002-5258-8761]{Abdurro'uf}
\affiliation{Department of Astronomy, Indiana University, 727 East Third Street, Bloomington, IN 47405, USA}
\email{fnuabdur@iu.edu}

\author[0000-0002-7570-0824]{Hakim Atek}
\affiliation{Institut d'Astrophysique de Paris, CNRS, Sorbonne Universit\'e, 98bis Boulevard Arago, 75014, Paris, France}
\email{atek@iap.fr}

\author[0000-0001-5984-0395]{Maru\v{s}a Brada{\v c}}
\affiliation{University of Ljubljana, Department of Mathematics and Physics, Jadranska ulica 19, SI-1000 Ljubljana, Slovenia}
\affiliation{Department of Physics and Astronomy, University of California Davis, 1 Shields Avenue, Davis, CA 95616, USA}
\email{marusa@ucdavis.edu}

\author[0000-0002-7908-9284]{Larry D. Bradley}
\affiliation{Space Telescope Science Institute (STScI), 3700 San Martin Drive, Baltimore, MD 21218, USA}
\email{lbradley@stsci.edu}

\author[0000-0002-8651-9879]{Andrew J.\ Bunker}
\affiliation{Department of Physics, University of Oxford, Denys Wilkinson Building, Keble Road, Oxford OX1 3RH, UK}
\email{Andy.Bunker@physics.ox.ac.uk}

\author[0000-0001-5487-0392]{Viola Gelli}
\affiliation{Cosmic Dawn Center (DAWN), Denmark}
\affiliation{Niels Bohr Institute, University of Copenhagen, Jagtvej 128, DK-2200 Copenhagen N, Denmark}
\email{viola.gelli@nbi.ku.dk}

\author[0009-0009-3404-5673]{Novan Saputra Haryana}
\affiliation{Astronomical Institute, Tohoku University, 6-3, Aramaki, Aoba, Sendai, Miyagi 980-8578, Japan}
\email{novan.haryana@astr.tohoku.ac.jp}

\author[0000-0001-8587-218X]{Matthew J. Hayes}
\affiliation{Stockholm University, Department of Astronomy and Oskar Klein Centre for Cosmoparticle Physics, AlbaNova University Centre, SE-10691, Stockholm, Sweden}
\email{matthew.hayes@astro.su.se}

\author[0000-0003-3367-3415]{George Helou} 
\affiliation{IPAC, California Institute of Technology, MC 314-6, 1200 E. California Boulevard, Pasadena, CA 91125, USA}
\email{gxh@ipac.caltech.edu}

\author[0000-0003-4570-3159]{Nicha Leethochawalit}
\affiliation{National Astronomical Research Institute of Thailand (NARIT), Mae Rim, Chiang Mai, 50180, Thailand}
\email{nicha@narit.or.th}

\author[0009-0002-8965-1303]{Zhaoran Liu}
\affiliation{Astronomical Institute, Tohoku University, 6-3, Aramaki, Aoba, Sendai, Miyagi 980-8578, Japan}
\affiliation{MIT Kavli Institute for Astrophysics and Space Research, 70 Vassar Street, Cambridge, MA 02139, USA}
\email{zhaoran.liu@astr.tohoku.ac.jp}

\author[0000-0002-9946-4731]{Marc Rafelski}
\affiliation{Space Telescope Science Institute, 
3700 San Martin Drive, Baltimore, MD, 21218 USA}
\affiliation{Department of Physics and Astronomy, Johns Hopkins University, Baltimore, MD 21218,USA}
\email{mrafelski@stsci.edu}

\author[0000-0002-4140-1367]{Guido Roberts-Borsani}
\affiliation{Department of Physics \& Astronomy, University College London, London, WC1E 6BT, UK}
\email{g.robertsborsani@ucl.ac.uk}

\author[0000-0001-7016-5220]{Michael J. Rutkowski}
\affiliation{Minnesota State University, Mankato, Department of Physics and Astronomy, 141 Trafton Science Center N, Mankato, MN 56001, USA}
\email{michael.rutkowski@mnsu.edu}

\author[0000-0002-9136-8876]{Claudia Scarlata}
\affiliation{University of Minnesota, Twin Cities, 116 Church St SE, Minneapolis, MN 55455, USA}
\email{}

\author[0000-0001-9935-6047]{Massimo Stiavelli}
\affiliation{Space Telescope Science Institute, 3700 San Martin Drive, Baltimore, MD 21218, USA}
\email{mstiavel@stsci.edu}

\author[0009-0005-1487-7772]{Ryo A. Sutanto}
\affiliation{Astronomical Institute, Tohoku University, 6-3, Aramaki, Aoba, Sendai, Miyagi 980-8578, Japan}
\email{ryo.sutanto@astr.tohoku.ac.jp}

\author[0000-0002-7064-5424]{Harry I. Teplitz}
\affiliation{IPAC, Mail Code 314-6, California Institute of Technology, 1200 E. California Blvd., Pasadena CA, 91125, USA}
\email{hit@ipac.caltech.edu}

\author[0000-0002-8460-0390]{T. Treu}
\affiliation{Department of Physics and Astronomy, University of California, Los Angeles, 430 Portola Plaza, Los Angeles, CA 90095, USA}
\email{tt@astro.ucla.edu}

\author[0000-0001-9391-305X]{M. Trenti}
\affiliation{School of Physics, The University of Melbourne, VIC 3010, Australia}
\email{michele.trenti@unimelb.edu.au}

\author[0000-0003-0980-1499]{Benedetta Vulcani}
\affiliation{INAF -- Osservatorio Astronomico di Padova, Vicolo Osservatorio 5, 35122 Padova, Italy}
\email{benedetta.vulcani@inaf.it}

\author[0000-0002-9373-3865]{Xin Wang}
\affiliation{School of Astronomy and Space Science, University of Chinese Academy of Sciences (UCAS), Beijing 100049, China}
\affiliation{National Astronomical Observatories, Chinese Academy of Sciences, Beijing 100101, China}
\affiliation{Institute for Frontiers in Astronomy and Astrophysics, Beijing Normal University, Beijing 102206, China}
\email{xwang@ucas.ac.cn}


\begin{abstract}

We present photometric properties of 161 galaxy candidates at $z=7-14$ selected from the second data release (DR2) of BEACON, a JWST Cycle 2 pure-parallel NIRCam imaging program. Carefully selected from 36 independent pointings (corresponding to $\sim350$\,arcmin$^2$ sky coverage), and hence with reduced cosmic variance, our galaxy candidates provide an unbiased sample for investigating galaxy properties over a wide range of environments. We measure the physical properties, including UV continuum slope ($\beta_{\rm UV}$), stellar mass ($M_*$), star formation rate (SFR), and sizes. Our highest redshift galaxy candidate at $z=13.71\pm0.15$ has a remarkably bright UV luminosity of $M_{\rm UV}=-21.19\pm0.08$, making it the brightest galaxy at $z>12$ if spectroscopically confirmed. With an extremely blue UV slope, compact morphology, and high star formation rate surface density ($\Sigma_{\rm SFR}$), this candidate may have extremely low metallicity, high ionizing photon escape fraction, or contributions from an AGN. Among our multiple independent sightlines, we identify three fields of galaxy number overdensity with $>3\sigma$ significance. The properties of galaxies in various environments do not exhibit significant differences, implying either that accelerated galaxy evolution in overdense regions is not yet widespread at $z>7$, or that the current constraints are limited by sample size. Our simulations indicate that increasing the sample by an order of magnitude would allow such environmental trends to be robustly confirmed or ruled out, underscoring the importance of future pure-parallel observations. 

\end{abstract}


\section{Introduction} \label{sec:intro}
JWST has opened up a new era for high-redshift galaxy searches, extending the frontier of galaxy detection to the first 300~million years after the Big Bang ($z\sim14$) during the earliest stages of cosmic reionization. Photometric selections with NIRCam imaging have found a large population of luminous, potentially massive galaxy candidates at very high redshifts \citep[e.g.,][]{Castellano2022,naidu22lf,finkelstein23lf,harikane23lf,robertson23lf,robertson24lf,adams24lf,franco25} that was unanticipated by pre-JWST theoretical models and observations. A large fraction of these galaxy candidates were later confirmed by extensive spectroscopic follow-up observations \citep[e.g.,][]{cl22,harikane24lf,carniani25,naidu25,rb25b}, verifying the NIRCam photometric selection and confirming the excess of bright galaxies at $z>10$. Possible explanations of such an excess of bright galaxies at high redshifts include: \textit{i)} higher average luminosity of galaxies due to elevated star formation efficiency \citep[e.g.,][]{inayoshi22,dekel23} or a top-heavy initial mass function \citep[IMF; e.g.,][]{cueto24,hutter25}, \textit{ii)} more bursty/stochastic star formation than at lower redshifts, that causes more low-mass dark matter (DM) halos to host bright galaxies temporarily at certain times \citep[e.g.,][]{Mason2023,shen23,gelli24}, or \textit{iii)} extremely low dust attenuation in the UV \citep[e.g.,][]{Mason2023,ferrara23,fiore23,ziparo23,ferrara24,ferrara25a,ferrara25b}. However, it remains a challenge to choose between these scenarios with current observations.

Another main observational challenge of characterizing bright galaxies is that they tend to be highly clustered, making even large surveys strongly affected by cosmic variance. With most previous $z>7$ galaxy selections based on JWST data focusing on a limited number of legacy fields, it is unclear how representative they are of the entire universe. Several galaxy overdensities have been identified in these legacy fields \citep[e.g.][]{Castellano2023,helton24a,helton24,chen25,li25od,napolitano25,witten25} at high redshifs with elevated star formation rate (SFR), suggesting that our current understanding of galaxy formation in the first several hundred million years may be biased towards a small number of unique sites of unusually prompt strong star formation. To overcome such potential biases from cosmic variance, it is necessary to investigate in an unbiased way the general properties of galaxies in the early universe with multiple independent sight-lines that supplement the legacy fields.

Pure-parallel observations offer an ideal opportunity to conduct such studies. Since the early 2010s, several HST pure-parallel programs have successfully identified luminous ($M_{\rm UV}= -21\sim-23$) galaxy candidates at $z\sim8-11$ \citep[e.g.,][]{trenti11,bradley12,morishita20,RR20,morishita21,rb21b,rb25a,rr25}. Pure-parallel programs with JWST slitless spectroscopy also allow direct redshift confirmation of galaxies up to $z=9$ with strong emission lines \citep[e.g.,][]{malkan25,sun25}. JWST NIRCam, covering the rest-UV and optical wavelengths, now allows us to extend such attempts out to $z>10$, and study galaxy properties and luminosity functions over a wide range of environments at cosmic dawn \citep[see also][for an overview of the Cycle 1 NIRCam pure-parallel program, PANORAMIC]{williams25}.

Here we present analysis of 161 galaxy candidates at $z>7.3$ selected from the data release 2 (DR2) of the Bias-free Extragalactic Analysis for Cosmic Origins with NIRCam (BEACON) program \footnote{The data products of BEACON DR2 will be available at \url{https://beacon-jwst.github.io/data.html}.} \citep[GO-3990; PI: T. Morishita; Co-PIs: C. Mason, T. Treu, M. Trenti;][]{morishita25}, a Cycle 2 NIRCam pure-parallel imaging program optimized for detections of $z>7.3$ Lyman break galaxies over a total of 69 independent pointings (equivalent to $670$\,arcmin$^2$). In this work, we make use of 36 pointings (covering $\sim350$\,arcmin$^2$) from BEACON data that are covered by $\geq6$ filters where robust selections of $z>7.3$ Lyman break galaxy candidates can be performed. To investigate whether the physical properties of galaxies differ between the bias-free pure-parallel fields and legacy fields, we carefully analyze the galaxy properties, including photometric redshift, of these galaxy candidates, and investigate their correlations with local galaxy density. The UV luminosity functions and clustering properties of the same galaxy sample will be presented in the accompanying paper (Kreilgaard et al., in preparation).

This paper is structured as follows. We describe the BEACON DR2 in Section \ref{sec:data}, followed by the $z\gtrsim7$ galaxy candidate selection in Section \ref{sec:sample}. We detail the spectral energy distribution (SED) fitting procedure in Section \ref{sec:sed}, and present the derived galaxy properties in Section \ref{sec:result_properties}. In Section \ref{sec:size}, we investigate the sizes of our sample through simple image analysis. Section \ref{sec:discuss} details a remarkably bright galaxy at $z=13.7$ and discusses how galaxy properties differ in various environments. Throughout the paper, we use AB magnitudes and the cosmological parameters of $\Omega_m=0.3$, and $\Omega_\Lambda=0.7$, $H_0 = 70$ $\mathrm{km s}^{-1}\mathrm{ Mpc}^{-1}$.

\section{Data} \label{sec:data}
Here we summarize the BEACON DR2 NIRCam imaging data and multiband photometric catalog construction. We refer the reader to \cite{morishita25} for the details of the survey design and data reduction. 

BEACON DR2 includes 69 independent pointings observed in parallel with primary observations, taken from 2023 October to 2024 December. Each pointing was observed with two to sixteen NIRCam medium/broad band filters, with the exposure times for each filter ranging from 408~s to 18940~s. To ensure a robust selection of $z\gtrsim7$ galaxy candidates, we limit our analyses to 36/69 pointings that are covered by six or more broad/medium band filters, resulting in a total sky coverage of 350~arcmin$^2$. 

The imaging data in these 36 pointings are reduced following the procedures outlined in \citet{morishita25}. Briefly, we retrieved the raw-level images from the Mikulski Archive for Space Telescopes (MAST) archive and reduced them with the official JWST pipeline, with additional steps including $1/f$-noise subtraction using {\tt bbpn}\footnote{\url{https://github.com/mtakahiro/bbpn}}, ``snowball" masking using {\tt Grizli} \citep{brammer22}, and additional cosmic-ray masking using {\tt lacosmic} \citep{vandokkum01,bradley23}. The final drizzled images, with a pixel scale set to $0.\!''0315$\,/\,pixel, are aligned to the IR-detection image (i.e., inverse-variance-weighted sum of F277W, F356W, and F444W).

We perform source detection and photometry using {\tt Source Extractor} \citep{1996A&AS..117..393B}. To ensure accurate color measurement, we first match the point spread functions (PSFs) of each filter image to that of the F444W image in each pointing. We generate PSFs in each filter with {\tt stpsf}\footnote{\url{https://github.com/spacetelescope/stpsf}}, which are then fed to {\tt pypher} \citep{boucaud16} to generate the convolution kernels for each filter. We run {\tt Source Extractor} in dual image mode, performing source detection in the detection images and photometry in each PSF-matched filter image. The key configuration parameters of {\tt Source Extractor} are set as follows: DETECT\_MINAREA 0.0081\,arcsec$^2$, DETECT\_THRESH=1.0, DEBLEND\_NTHRESH=64, DEBLEND\_MINCONT=0.0001, BACK\_SIZE=128, and BACK\_FILTSIZE=5. We measure the source fluxes within fixed apertures of $0.\!''16$ radii, and then scale them to the total fluxes with a single factor for each source, defined as ${f_{\rm auto, F444W}/f_{\rm aper, F444W}}$, where ${f_{\rm auto}}$ is the flux measured within elliptical Kron apertures with a 2.5 scaling factor \citep{1996A&AS..117..393B}. Finally, we correct for the Galactic dust reddening using the attenuation value retrieved for the coordinates of each field from NED \citep{schlegel98,schlafly11}, assuming the Milky Way reddening law \citep{cardelli89}.


\section{Sample Selection} \label{sec:sample}
From the multiband source catalog constructed in Section \ref{sec:data}, we select $z\gtrsim7$ galaxies via the Lyman break dropout method and constrain their redshift probability distribution, $p(z)$, with {\tt EAZY} \citep{brammer08}. We detail the selection procedure below.

We first apply the following color selection criteria in three redshift bins:

{\bf F090W-dropouts ($7.3\lesssim z \lesssim 9.7$)}
$$S/N_{\rm 150} > 4 $$
$$S/N_{\rm 090} < 2$$
$$z_{\rm set} = 6$$

{\bf F115W-dropouts ($9.7\lesssim z \lesssim13$)}
$$S/N_{\rm 200} > 4 $$
$$S/N_{\rm 115, 090} < 2$$
$$z_{\rm set} = 8$$

{\bf F150W-dropouts ($13\lesssim z \lesssim 18$)}
$$S/N_{\rm 277} > 4 $$
$$S/N_{\rm 150, 115, 090} < 2$$
$$z_{\rm set} = 10,$$
which require $2\sigma$ non-detections for filters that are bluer than the rest-frame 1216~\AA\ of the sources at the corresponding redshift ranges. To further ensure non-detections, we repeat the non-detection step with a smaller aperture, $r=0.\!''08$ ($\sim2.5$\,pixel). The $z_{\rm set}$ parameter 
is the redshift limit used for the $z_{\rm phot}$ cut (see below).
For typical noise levels, these selections correspond to Lyman breaks of 0.7--4.2 magnitudes. 

Next, we derive $p(z)$ of the color-selected sources with {\tt EAZY}. We adopt the template sets in \citet{hainline23}, with the addition of dusty emission line galaxy template from \citet{naidu22sed}. The \citet{hainline23} templates supplement the default {\tt EAZY} ``v1.3'' library with young, high specific star formation rate (sSFR) galaxy templates generated with {\tt fsps} \citep{conroy09fsps}, enabling a more comprehensive redshift estimate for our target young galaxies. We set the redshift range of fitting to $0<z<20$ with a step size of $0.01 (1+z)$. In each redshift bin considered, {\tt EAZY} combines all of the available templates together and applies the average \citet{madau95} IGM absorption.
The $\chi^2$ of the best fit in that redshift bin is recorded as $\chi^2(z)$ which is output from the program. We did not adopt any apparent magnitude priors, as the galaxy apparent magnitude - redshift relation at $z>7$ is currently not well constrained. To prevent bright fluxes from dominating the fits, we set an error floor on the photometry of 5\%. We also use the {\tt EAZY} template error file “{\tt TEMPLATE\_ERROR.eazy\_v1.0}” to account for uncertainties in the templates as a function of wavelength.

We run {\tt EAZY} with the above settings and obtain the redshift probability distribution, $p(z)$, of each source from $\chi^2(z)$, assuming a uniform redshift prior: $p(z)=\exp[-\chi^2(z)/2]$, which is normalized such that $\int p(z)dz=1.0$. We then apply the $z_{\rm phot}$ cut, requiring $p(z>z_{\rm set}) > 0.8$, i.e., the total probability that each source is at $z>z_{\rm set}$ must be greater than 80\%, where $z_{\rm set}$ is defined above for each redshift bin. The combined color selection and $p(z)$ cut ensure the selected sources have redshifts consistent with the redshift range of the dropout selections. With these procedures, we obtain 266 sources.

From here, we conduct visual inspections on the 266 sources, rejecting data artifacts consisting of extended diffraction spikes from stars, bad pixels, and hot pixels caused by cosmic rays in any of the available filters that may affect the photometry. We are left with 166 candidates after visual inspection, including 151, 14, and one sources with F090W-dropouts, F115W-dropouts, and F150W-dropouts, respectively. These sources represent our final sample of $z\gtrsim 7$ galaxy candidates.

One main source of contamination for high-$z$ galaxy selection is foreground low-mass stars (T-, L-, M-type stars, i.e., brown dwarfs), which have similar spectral features to high-$z$ galaxies such as the color break at observed $\sim1\,\mu$m \citep{morishita21,hainline24dwarf,greene24lrd}. To assess such contamination, we use \texttt{EAZY} to fit the dwarf templates from \texttt{SPEX} library \citep{rayner03} to our 166 galaxy candidates at $z>7$, fixing the redshifts to zero. We find that for all 166 objects, the best-fit $z>7$ galaxy SEDs have lower $\chi^2$ than the best-fit dwarf SEDs, suggesting that our high-$z$ galaxy sample is unlikely to contain many foreground brown dwarfs.   

\begin{figure}[ht!]
\begin{center}
\includegraphics[scale=0.67]{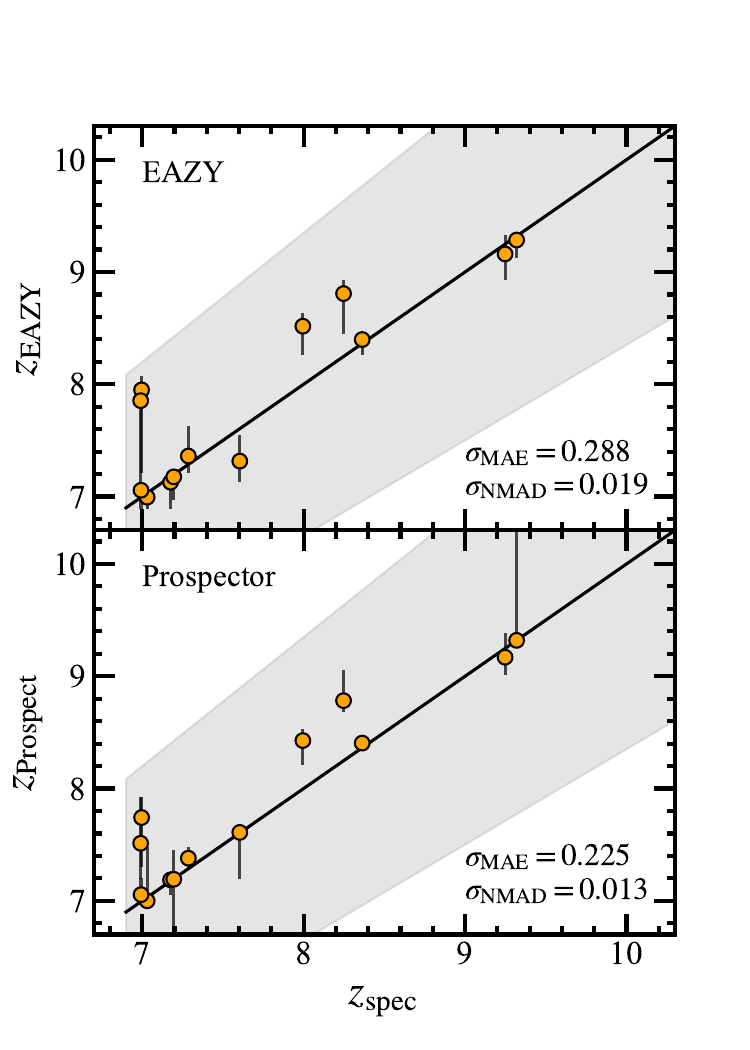}
\end{center}
\caption{Top: Comparison between \texttt{EAZY} photometric redshift and spectroscopic redshift in our sample. The black solid line indicates one-to-one relation, while the grey shaded region shows the $|z_{\rm phot}-z_{\rm spec}|/(1+z_{\rm spec})<0.15$. Bottom: similar to the top panel, but for \texttt{Prospector} photometric redshift.
}\label{fig:zcompare}
\end{figure}

Next, we examine the accuracy of the $p(z)$ derived by \texttt{EAZY}. We cross-match our 166 candidates with public JWST spectroscopy, using the spectroscopic catalog from Dawn JWST Archive (DJA) {\tt v4.4}. We find 14/166 objects in our final catalog of $z\gtrsim 7$ galaxy candidates have reliable $z_{\rm spec}$ measurements in the DJA catalog. All of these 14 galaxies have \texttt{zgrade=3} in the DJA catalog, with $z_{\rm spec}$ measured by strong emission lines in the PRISM or medium resolution spectra. In the top panel of Figure \ref{fig:zcompare}, we compare the EAZY photometric redshift ($z_{\rm EAZY}$) with $z_{\rm spec}$ of these 14 objects. All 14 objects are confirmed to be $z>7$ galaxies, and their $z_{\rm EAZY}$ is consistent with $z_{\rm spec}$ within the errors of $|z_{\rm phot} - z_{\rm spec}|/(1+z_{\rm spec}) < 0.15$, indicating that our sample selection effectively excludes low-$z$ interlopers. We further quantify the quality of $z_{\rm EAZY}$ with mean absolute error (MAE) and median absolute deviation (NMAD), which are defined in the following equations:
\begin{equation}
    \sigma_{\rm MAE} = \frac{\Sigma_{i=0}^N|\Delta z_i|}{N},
\end{equation}
\begin{equation}
    \sigma_{\rm NMAD} = 1.48\times{\rm median}\left\{\frac{|\Delta z - {\rm median}(\Delta z)|}{1+z_{\rm spec}}\right\},    
\end{equation}
where $\Delta z= z_{\rm EAZY}-z_{\rm spec}$. We obtain $\sigma_{\rm MAE}= 0.288$ and $\sigma_{\rm NMAD}=0.019$, indicating the good quality of \texttt{EAZY} photometric redshifts. We fix the redshift of these galaxies to $z_{\rm spec}$ when conducting SED fitting in the following section.

\begin{deluxetable*}{ccc}
\tablewidth{0pt}
\tablecaption{Summary of the 10 free parameters used for the SED fitting with {\tt Prospector} \label{tab:prospector_params}}
\tablehead{
\colhead{Parameter} & \colhead{Description} & \colhead{Prior}
}
\startdata
$z_{\rm phot}$ & Redshift & $p(z)$ obtained from {\tt Eazy} (Section \ref{sec:data}) \\
 &  & or fixed to $z_{\rm spec}$ \\
$\log(Z/Z_\odot)$ & Stellar metallicity & tophat: min=-2.0, max=0.0 \\
$\log(M_*/M_\odot)$ & Total stellar mass formed & tophat: min=6.0, max=11.0 \\
SFH & flexible SFH: ratio of the SFRs in adjacent time bins of the five-bin & student T: $\mu$=0.0, $\tau$=0.3, $\nu$=2.0 \\
& SFH (four parameters in total); & \\
$\hat{\tau}_{\rm dust,2}$ & diffuse dust optical depth & tophat: min=0.0, max=2.0 \\
$\log U$ & ionization parameter for nebular emission & tophat: min=-3.5, max=-1.0 \\
\enddata
\end{deluxetable*}

\begin{figure*}[ht!]
\begin{center}
\includegraphics[scale=0.72]{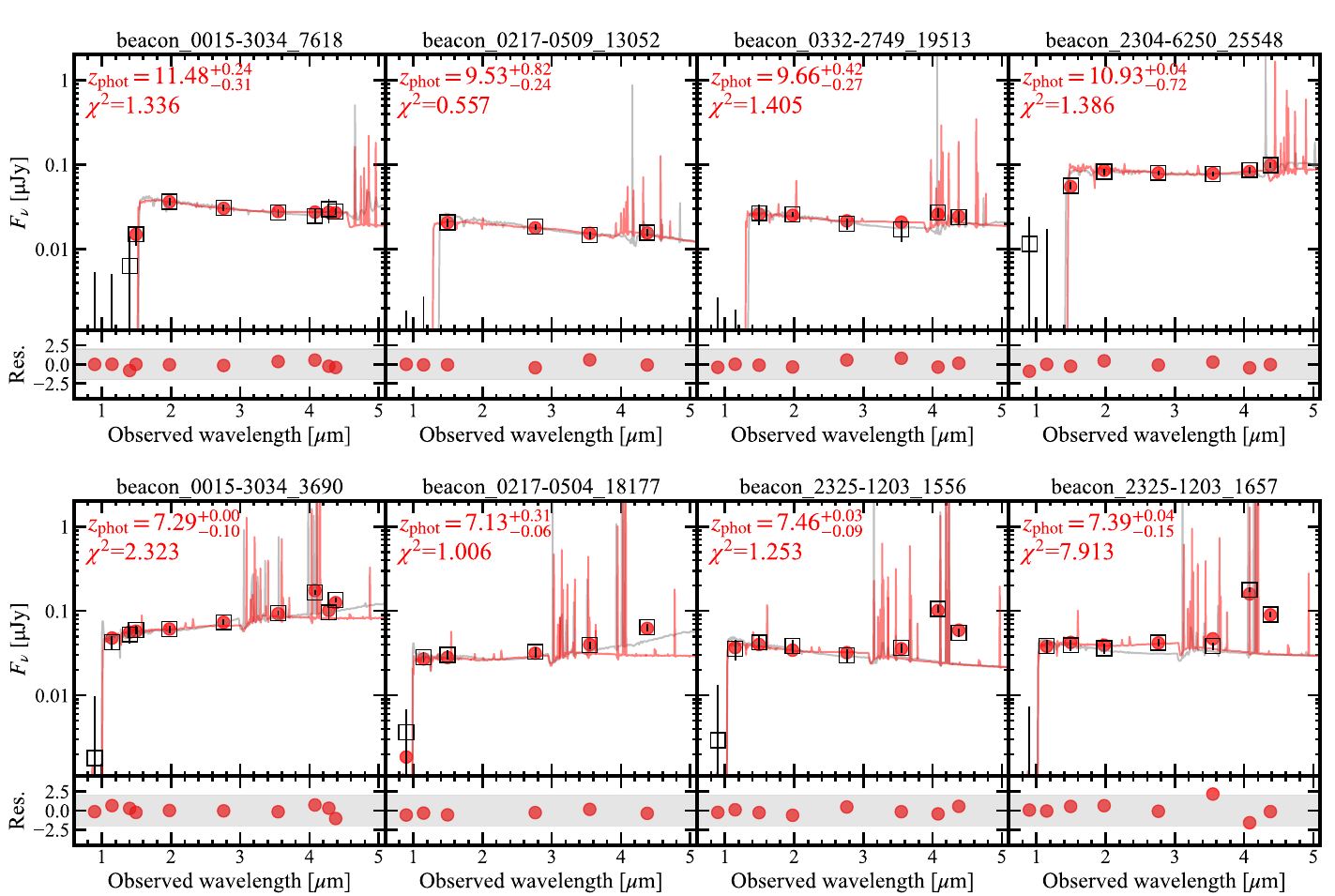}
\end{center}
\caption{Examples of our SED fitting results. In the top of each panel, we show the best-fit SEDs from \texttt{EAZY} and \texttt{Prospector} with grey and red curves, respectively. The red solid circles and black open squares indicate the photometry predicted from the best-fit SEDs and the observed photometry, respectively. The normalized residuals are presented in bottom of each panel, with the horizontal shaded region showing the $\pm2\sigma$ interval.
}\label{fig:sed}
\end{figure*}

\section{SED fitting}\label{sec:sed}
To derive the physical properties of photometric redshift ($z_{\rm phot}$), $M_*$, and SFR, for our $z\gtrsim 7$ galaxy candidates, we conduct SED fitting with \texttt{Prospector} \citep{prospect}. The model spectra are generated from the Flexible Stellar Population Synthesis \citep[FSPS;][]{conroy09fsps,cg10} package with the Modules for Experiments in Stellar Astrophysics Isochrones and Stellar Tracks \citep[MIST;][]{choi16} that includes the boost of the ionizing flux production of massive stars \citep{choi17}. The redshift is set to have the range of $0<z<20$, with the prior $p(z)$ obtained from {\tt EAZY} fitting in Section \ref{sec:sample} for each candidate. 
We assume a \citet{chab03} IMF, and set the priors of $M_*$ to be a tophat distribution with $6.0\leq\log(M_*/M_\odot)\leq12.0$. We assume a constant galaxy stellar metallicity $Z$ with a tophat prior of $-2\leq\log(Z/Z_\odot)\leq0$. For the SF history, we adopt a non-parametric continuity SFH \citep{leja19} with $N_{\rm SFH}$ time bins, where the SFR in each bin is constant. We adopt $N_{\rm SFH} = 5$, with the first time bin fixed at $0-10$~Myr of look-back time to capture the most recent variation in the SFH, while the remaining four bins are spaced equally in logarithm between the look-back time of 10~Myr and at $z=20$.  
We model the dust attenuation with the SMC dust extinction law \citep{gordon03}, setting the attenuation $A_V$ to have a tophat prior with the range of $0.0\leq A_V\leq2.0$. For the nebular attenuation, we fix the gas-phase metallicity to $Z$. 
We assume a tophat prior for the ionization parameter ($U$) with $-3.5\leq\log U\leq -1.0$.
The model spectra are then attenuated with the redshift-dependent \citet{madau95} average intergalactic medium (IGM) attenuation model. The summary of free parameters and their priors is listed in Table \ref{tab:prospector_params}.

\begin{figure*}[htb!]
\begin{center}
\includegraphics[scale=0.64]{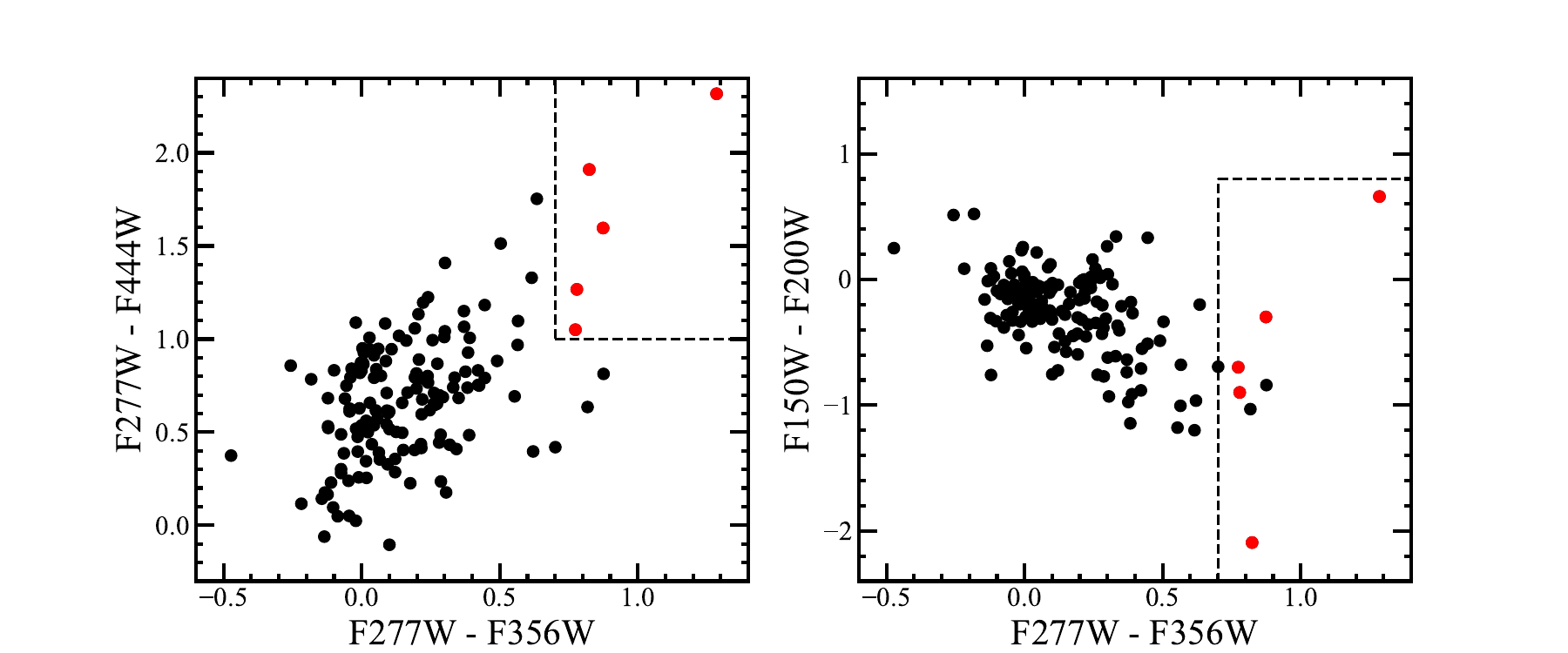}
\end{center}
\caption{Rest-frame UV and optical colors of our galaxy candidates at $z>7$ selected in Section \ref{sec:sample}. Red data points denote the five objects that are excluded due to their red optical color. The black dashed lines indicates the ``little red dot'' (LRD) selection criteria from \citet{greene24lrd} and \citet{hainline25lrd}.
}\label{fig:lrd}
\end{figure*}

We fit our SED model to the photometric data (Section \ref{sec:data}), obtaining the posterior SEDs and the posterior distributions of the free parameters listed in Table \ref{tab:prospector_params} using the Markov Chain Monte Carlo method implemented by \texttt{EMCEE}. We adopt the maximum-likelihood value as the best-fit estimate and quote uncertainties based on the inner 68th percentile of the posterior distribution.
Figure \ref{fig:sed} shows  examples of our best-fit SED models. Upon visual inspection on the SED fitting results, we find that most of our objects are well fitted by our SED models, with the residuals between models and observed photometry smaller than $2\sigma$. However, there are 5/166 objects whose rest-frame optical color is relatively red and the best-fit SED models significantly deviate from the photometric data. We examine their rest-frame UV and optical colors in Figure \ref{fig:lrd}, finding that the color of these six candidates are consistent with that of compact, red sources, i.e. ``little red dots'' (LRD), at $z>6$ found in multiple JWST studies \citep[e.g.,][]{akins23,furtak23lrd,kocevski23lrd,barro24lrd,greene24lrd,hainline25lrd}. The compactness of these five objects, defined as the flux ratio between the $0.\!''5$\ and $0.\!''2$\,diameter apertures in the F444W filter, ranges from $1.5-1.8$, which is also similar to LRDs \citep{greene24lrd,hainline25lrd}. 
Although the physical nature of LRDs remains unclear, a large fraction of LRDs are found to feature AGN activity and SEDs that cannot be fitted by typical stellar templates. To ensure the purity of our sample and consistency of our analyses, we exclude these five objects from the following analyses, resulting in a total of 161 objects in our sample.

As discussed earlier in this section, we use the $p(z)$ from \texttt{EAZY} as the priors of redshifts in \texttt{Prospector} SED fitting when $z_{\rm spec}$ is not available. Here we compare the performance of photometric redshift determination of \texttt{Prospector} and \texttt{EAZY} using the 14 galaxies at $z>7$ with confirmed $z_{\rm spec}$ (Section \ref{sec:data}). To perform such comparison, we run our standard SED fitting procedures mentioned earlier in this section on these 14 objects instead of fixing their redshifts to $z_{\rm spec}$. The resulting photometric redshifts ($z_{\rm Prospect}$) are compared with $z_{\rm spec}$ in the bottom panel of Figure \ref{fig:zcompare}. The $z_{\rm Prospect}$ of these 14 objects are in good agreement with their $z_{\rm spec}$, with the $\sigma_{\rm MAE}=0.225$ and $\sigma_{\rm NMAD}=0.013$ statistics substantially improved compared with the \texttt{EAZY} fitting results.

\begin{figure*}[ht!]
\begin{center}
\includegraphics[scale=0.74]{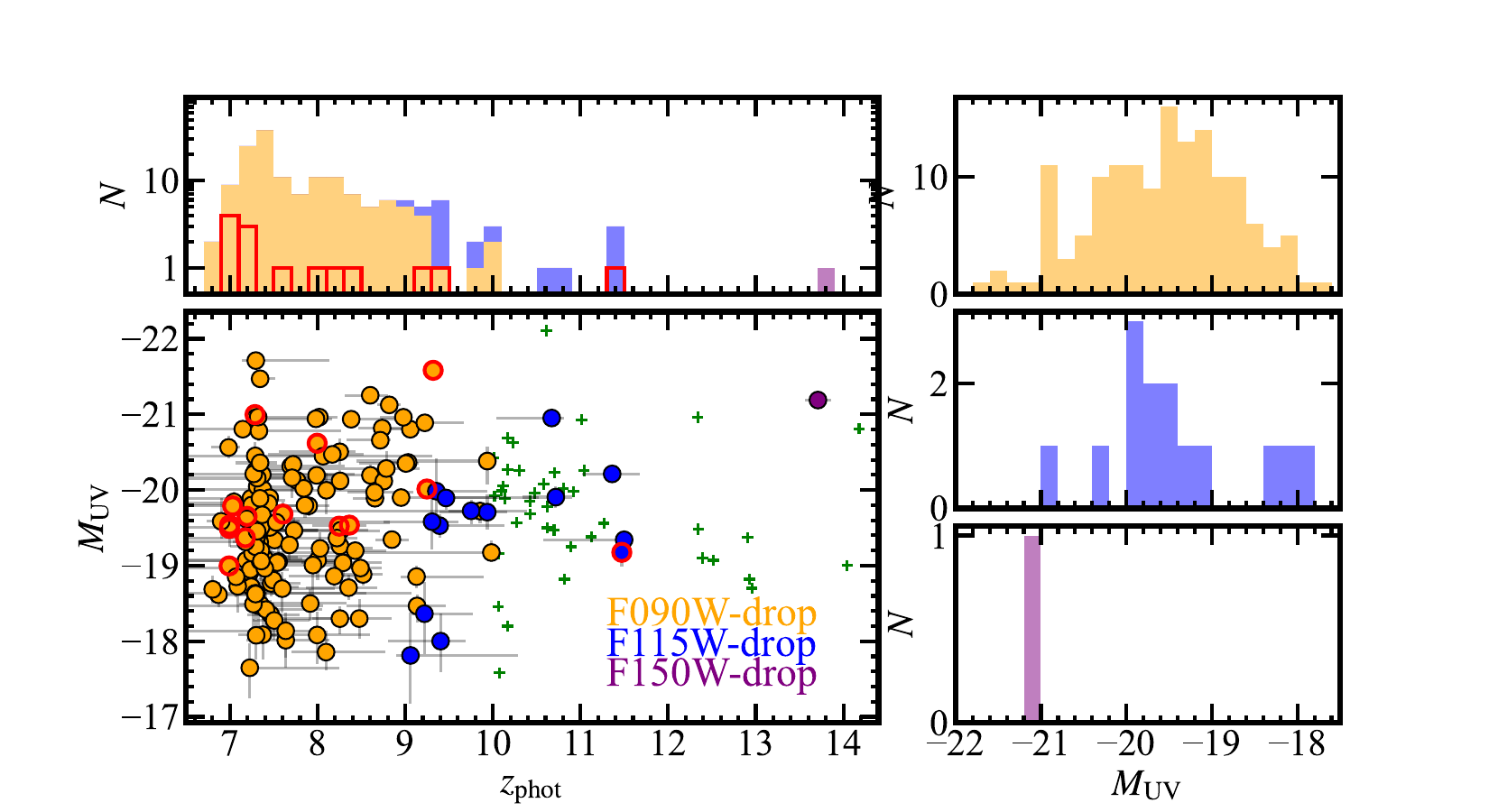}
\end{center}
\caption{Left: histogram of $z_{\rm phot}$ distribution (top) and $M_{\rm UV}$ versus $z_{\rm phot}$ (bottom) for our sample. Objects with spectroscopic confirmation are marked with open red symbols and the red histogram. We also plot other spectroscopically confirmed galaxies at $z>10$ in the literature with green data points. Right: histograms of $M_{\rm UV}$ distributions for the F090 (top), F115 (middle), and F150W (bottom) -dropout galaxies. 
}\label{fig:z_muv}
\end{figure*}

\section{Galaxy Properties} \label{sec:result_properties}
In this section, we present the properties of our 161 galaxy candidates derived from multiband photometry and SED fitting (Section \ref{sec:sed}).
\subsection{UV Continuum Properties} \label{subsec:uv}
We measure UV continuum properties, including the absolute UV magnitude ($M_{\rm UV}$) and the UV continuum slopes ($\beta_{\rm UV}$) by fitting a power law (i.e., $f_\lambda\propto\lambda^\beta$) to the NIRCam photometry covering the rest-frame UV. The filter sets adopted for UV continuum measurements are based on the $z_{\rm phot}$ obtained in Section \ref{sec:sed} and are selected to avoid potential Ly$\alpha$ contamination. For objects with $7.3\lesssim z_{\rm phot}\lesssim 9.7$ (F090W-dropouts), we use all the available filters including F150W, F200W, and F277W. For objects with $9.7\lesssim z_{\rm phot}\lesssim 13$ (F115W-dropouts), we use all the available filters including F200W, F277W, and F356W. For all 161 objects in our final sample, there are at least two filters used for UV continuum measurements. For each galaxy, we measure $M_{\rm UV}$ and $\beta_{\rm UV}$ with a Monte Carlo simulation, randomly drawing 200 realizations based on the measured values and errors of photometry and $z_{\rm phot}$. Figure \ref{fig:z_muv} shows the distribution of our galaxy candidates in the redshift-$M_{\rm UV}$ plane. Our sample covers a wide range of UV magnitude, spanning $M_{\rm UV}=-17\sim-22$. Noticeably, at $z=13.7$, our single F150W-dropout galaxy has a bright UV luminosity of $M_{\rm UV}=-21.2$, more luminous than any spectroscopically confirmed galaxy currently known at $z>12$. The detailed physical properties of this object will be discussed in \ref{subsec:z14}.



\begin{figure*}[ht!]
\begin{center}
\includegraphics[scale=0.7]{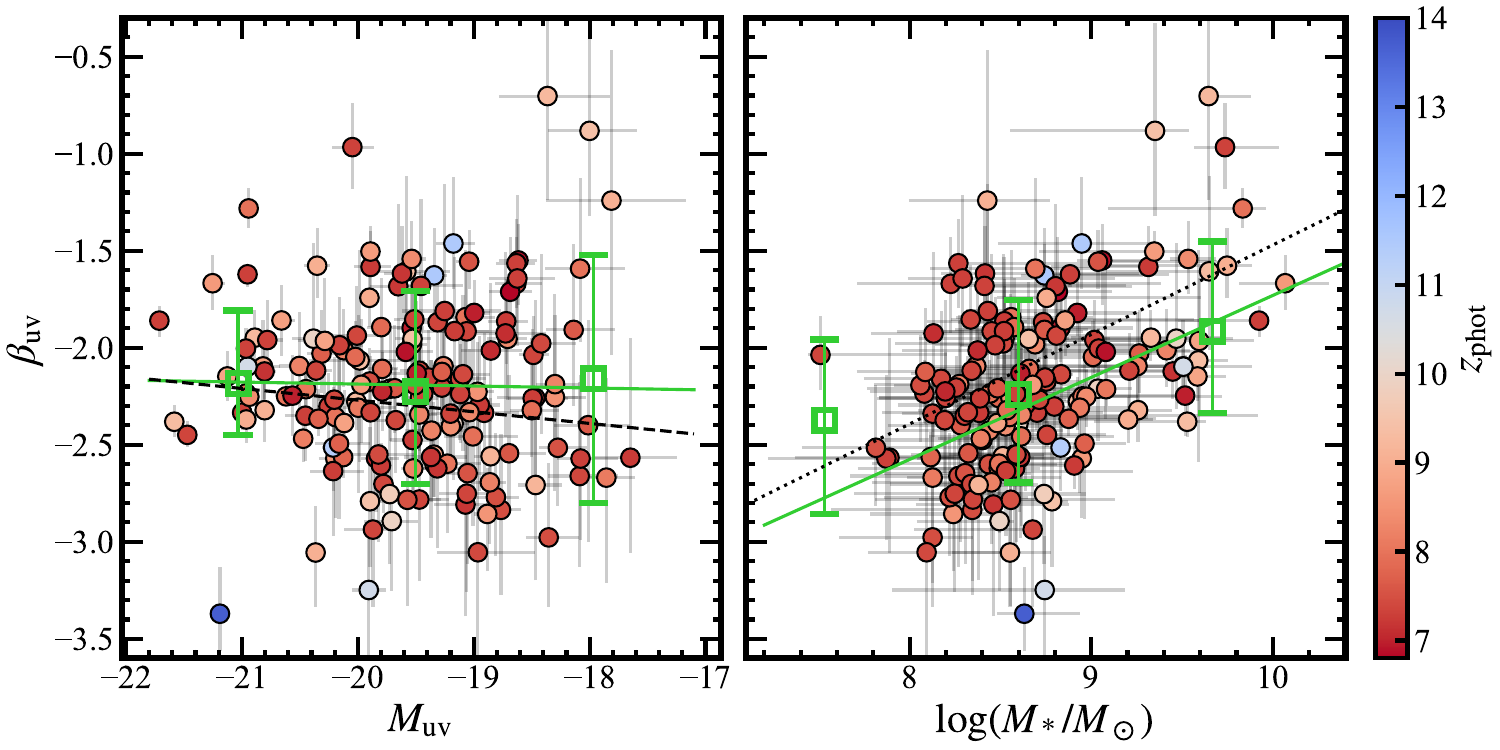}
\end{center}
\caption{Left: $\beta_{\rm UV}$ versus $M_{\rm UV}$ for our sample of 161 $z\geq7$ galaxies, where the data points are color-coded by redshift. The median $\beta_{\rm UV}$ and its inner 68 percentile distribution in different $M_*$ bins are denoted with green open squares and error bars. The green solid line indicates our best-fit $M_{\rm UV}-\beta_{\rm UV}$ linear relation. The black dashed line shows the $M_{\rm UV}-\beta_{\rm UV}$ relation from \citet{topping24}. Right: same as the left panel, but for $\beta_{\rm UV}$ versus $M_*$. The black dotted line shows the $M_*-\beta_{\rm UV}$ relation from \citet{finkelstein12}.
}\label{fig:beta}
\end{figure*}

In the left panel of Figure \ref{fig:beta}, we show the $M_{UV}-\beta_{\rm UV}$ relation of our sample, together with the median and inner 68 percentiles $\beta_{\rm UV}$ in different $M_{\rm UV}$ bins. The $\beta_{\rm UV}$ of our sample has $(16,50,84)$-percentile values of $(-2.61, -2.21, -1.75)$, with the majority ($\gtrsim70\%$) located at $\beta_{\rm UV} < -2.0$. There are four galaxies in our sample with extremely blue UV continuum slopes characterized by $\beta_{\rm UV} < -2.6$ even accounting for their measurement uncertainties. Such extremely blue galaxies, which have been reported to have low $M_*$ and high sSFR \citep[e.g.,][]{topping24,yanagisawa24}, are suspected to have large ionizing photon escape fractions ($f_{\rm esc}$). Due to the small number of such galaxies in our sample, we do not find strong correlation between extremely blue $\beta_{\rm UV}$ and $M_*$ or sSFR.

To investigate this (non-)correlation,
we fit our measurements with a linear relation $\beta_{\rm UV} = \frac{{\rm d}\beta}{{\rm d}M_{\rm UV}}M_{UV}^{-19}+\beta_0$ with an intrinsic scatter of $\sigma_{\beta,{\rm UV}}$. Here $M_{UV}^{-19}\equiv M_{\rm UV}+19$, and $\beta_0$ represents the $\beta_{\rm UV}$ at $M_{\rm UV}=-19$. We obtain the best-fit parameters ($\frac{{\rm d}\beta}{{\rm d}M_{\rm UV}}$, $\beta_0$, $\sigma_{\beta,{\rm UV}}$) with {\tt emcee}, incorporating the measurement uncertainties of $\beta_{\rm UV}$ and $M_{\rm UV}$. The derived relation, which is shown with the green solid line in the left panel of Figure \ref{fig:beta}, has the functional form of $\beta_{\rm UV} = -0.01_{-0.03}^{+0.04}\,M_{UV}^{-19}-2.19_{-0.03}^{+0.04}$ with an intrinsic scatter of $\sigma_{\beta,{\rm UV}} = 0.29_{-0.02}^{+0.02}$. The best-fit slope, $\frac{{\rm d}\beta_{\rm UV}}{{\rm d}M_{\rm UV}} = -0.01_{-0.03}^{+0.04}$, indicates that there is no correlation between $\beta_{\rm UV}$ and $M_{\rm UV}$, and is in agreement with the results $\frac{{\rm d}\beta_{\rm UV}}{{\rm d}M_{\rm UV}} = -0.06_{-0.05}^{+0.05}$ obtained from the JADES survey by \citet{topping24} 
at $z>8$. Similarly, \citet{franco25} also found a flat slope of $\frac{{\rm d}\beta_{\rm UV}}{{\rm d}M_{\rm UV}} = -0.03_{-0.03}^{+0.03}$. Our best-fit $\beta_0=-2.19_{-0.03}^{+0.04}$ is slightly redder than the value $-2.33_{-0.05}^{+0.05}$ reported in \citet{topping24}, but is within the error when taking into account the intrinsic scatter $\sigma_{\beta,{\rm UV}} = 0.31_{-0.02}^{+0.02}$.  

For the $M_*-\beta_{\rm UV}$ relation, we conduct similar analyses and present our results in the right panel of Figure \ref{fig:beta}. The $M_*-\beta_{\rm UV}$ relation of our sample is best described by a linear function of $\beta_{\rm UV} = 0.42_{-0.05}^{+0.04}\log(M_*/M_\odot)-5.96_{-0.34}^{+0.37}$ with an intrinsic scatter of $\sigma_{\beta,{\rm UV}} = 0.17_{-0.02}^{+0.02}$. The positive correlation is significant at the $>3\sigma$ level, consistent with previous studies at $z=4-8$ \citep{finkelstein12} and $z=9-11$ \citep{tacchella22}, indicates that more massive galaxies are redder in rest-UV color, and hence are likely more dusty or composed of older stellar populations. 

Overall, our UV continuum analyses suggest that the $M_{\rm UV}-\beta_{\rm UV}$ and $M_*-\beta_{\rm UV}$ relations of our sample do not deviate from those previously reported in legacy fields at the similar redshifts.

\subsection{Galaxy Main Sequence}
We derive the SFR of our galaxies based on the rest-frame UV luminosity ($L_{\rm UV}$) inferred from photometry. The UV luminosity is corrected for the dust attenuation ($A_{1600}$) that is estimated from the $\beta_{\rm UV}$ and the following relation \citep{meurer99}:
\begin{equation}
    A_{\rm UV} = 4.43 + 1.99\beta_{\rm UV}.
\end{equation}
The attenuation corrected $L_{\rm UV}$ is then converted to the UV SFR \citep{kennicutt98}:
\begin{equation}
    {\rm SFR}_{\rm UV}~[M_\odot~{\rm yr}^{-1}] = 1.4\times10^{-28} L_{\rm UV}~ [\mathrm{erg~s^{-1}~Hz^{-1}}],
\end{equation}
which is multiplied by 0.63 to account for the conversion from \citet{salpeter} IMF to \citet{chab03} IMF \citep{madau14}. We adopt the UV-based SFR measurements instead of directly using the SFR from SED fitting results because it is difficult to distinguish the detailed star forming history and other effects (e.g., dust attenuation, metallicity) of galaxies at $z>7$ with the current observational datasets \citep[e.g.,][]{tacchella22}.

\begin{figure}[ht!]
\begin{center}
\includegraphics[scale=0.57]{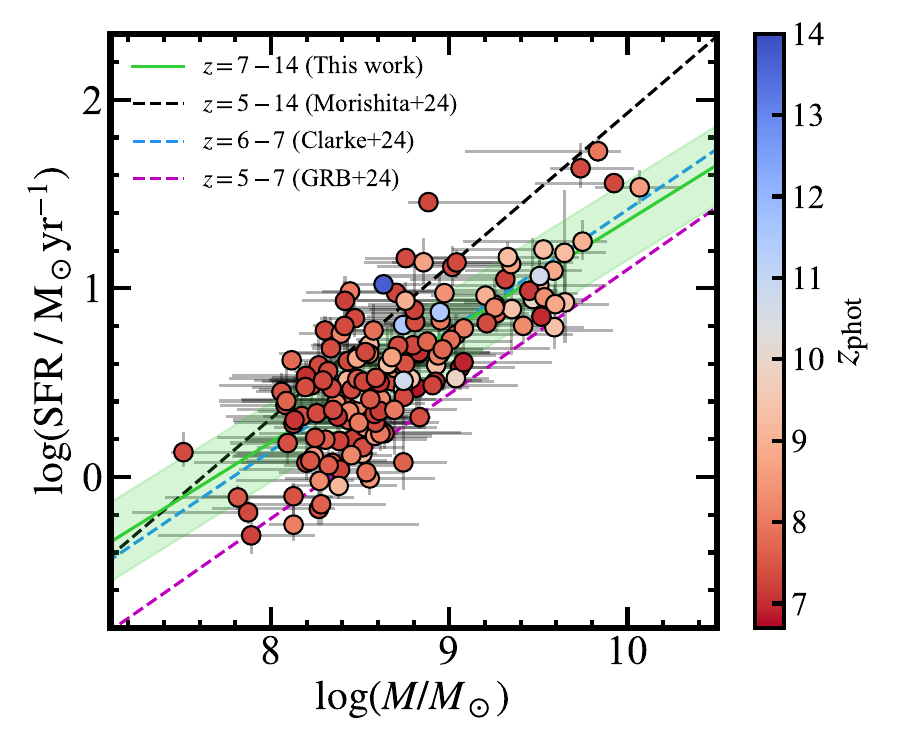}
\end{center}
\caption{The $M_*-$SFR relation of our galaxy sample. Data points are color-coded in the same manner as Figure \ref{fig:beta}. The green line and shaded regions indicate the best-fit SFR$= \alpha\log(M_*/M_0)+\beta$ relation and the intrinsic scatter, respectively. The black, blue, and magenta dashed lines denote previous results from \citet{morishita24}, \citet{clarke25}, and \citet{rb24}, respectively, at the similar redshift.
}\label{fig:ms}
\end{figure}

We present the galaxy main sequence ($M_*-$SFR relation) of our sample in Figure \ref{fig:ms}. We fit a line to the observed main sequence with log(SFR)$= \alpha\log(M_*/M_0)+\beta$ with an intrinsic scatter of $\sigma_{\rm ms}$, where $M_0=10^8M_\odot$. Our best-fit result has a slope of $\alpha=0.59_{-0.04}^{+0.03}$, a normalizaton of $\beta=0.18_{-0.03}^{+0.03}$, and an intrinsic scatter of $\sigma_{\rm ms} = 0.21_{-0.02}^{+0.02}$. In Figure \ref{fig:ms}, we also compare our result with other studies based on JWST-selected galaxies at the similar redshift. Overall, our best-fit main sequence is located between \citet{morishita24} and \citet{rb24}, and is in good agreement with \citet{clarke25}. 

\section{Galaxy sizes}\label{sec:size}
We estimate the sizes of galaxies with {\tt Galfit} \citep{peng02,peng10}, assuming a single simple surface brightness profile. Following \citet{morishita24}, we fix the S\'ersic index $n$ to 1, i.e., a pure exponential law. For each galaxy, we first generate image cutouts ($41 \times 41$ pixels in size, corresponding to ) of the non-PSF-matched science, rms, and segmentation maps. We then mask out the neighboring sources using the segmentation map. We construct the empirical PSF models by selecting and stacking bright stars in the same field. For each galaxy, we perform the fit in the filter covering the rest-frame UV wavelengths, i.e., F150W(F200W) for F090W(F115W)-dropouts, obtaining the best-fit model and the corresponding effective radius ($R_e$), which is defined as the radius where $50\%$ of the total flux is covered. 

\begin{figure}[ht!]
\begin{center}
\includegraphics[scale=0.4]{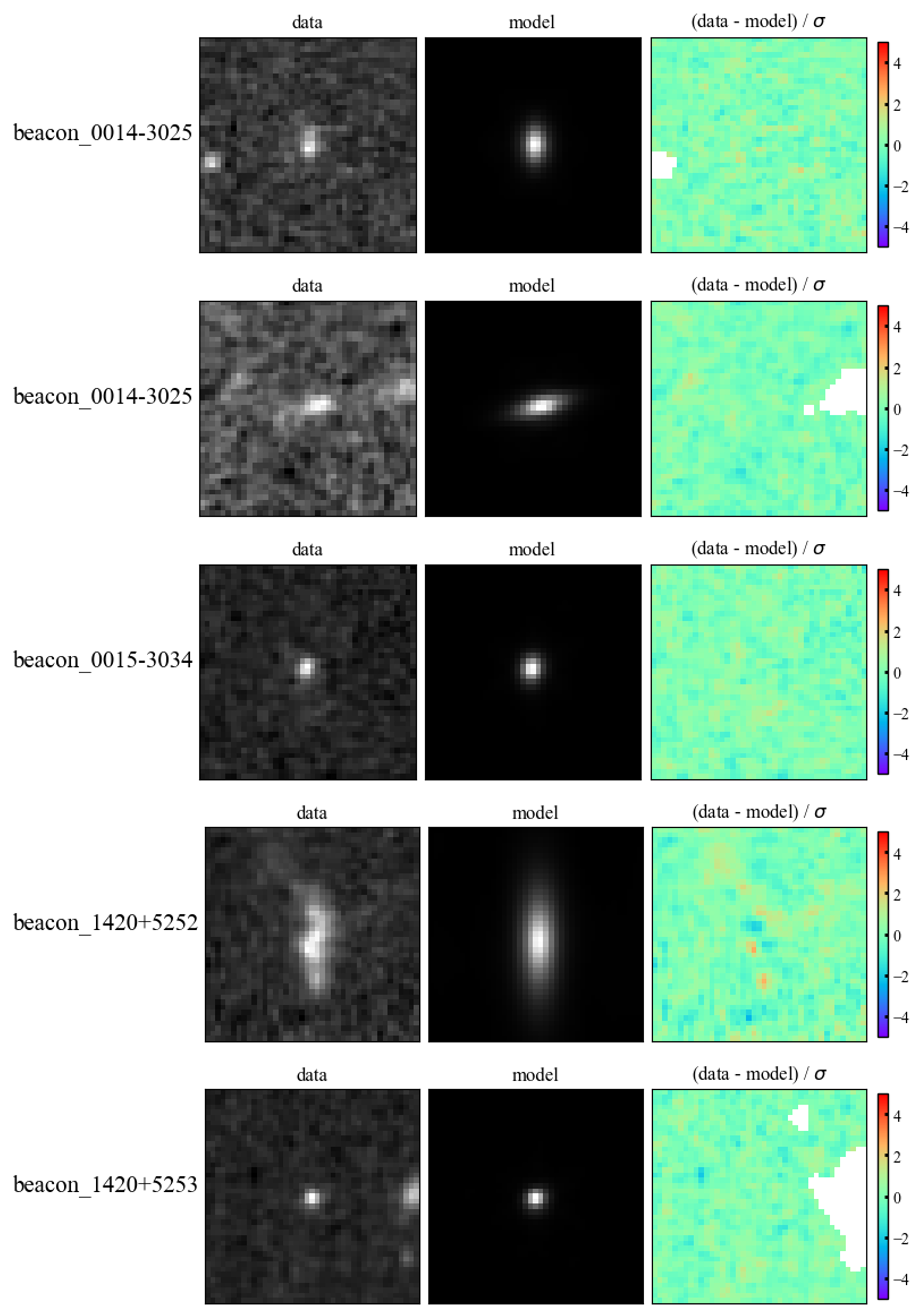}
\end{center}
\caption{Examples of \texttt{galfit} fitting results. The left, middle, and right columns show the cutout images of galaxies, best-fit S\'ersic profiles, and normalized residuals, respectively. Other sources are masked out in the last column.
}\label{fig:galfit}
\end{figure}

After obtaining the best-fit results for each galaxy, we conduct visual inspections on the residual images, identifying 13 objects with significant residuals, e.g., objects with multiple clumps or interacting with neighboring galaxies. These 13 objects are excluded from the following galaxy size analyses.

\begin{figure}[ht!]
\begin{center}
\includegraphics[scale=0.65]{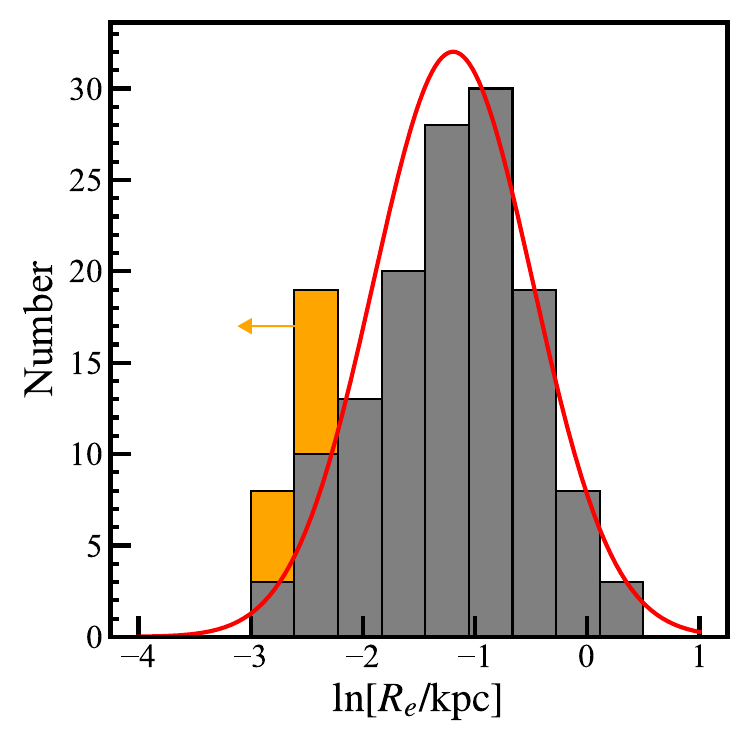}
\end{center}
\caption{Histogram of $R_e$ in natural logarithm scale. Measurements with upper limits are indicated with orange color. The red curve represents the best-fit log-normal distribution (see Section \ref{sec:size}).
}\label{fig:size_hist}
\end{figure}

Figure \ref{fig:galfit} shows examples of our {\tt Galfit} fitting results. Overall, there are 14 galaxies with upper limit of $R_e$=0.5\,pix ($\sim 57$\,pc at z=7). The sizes of our $z>7$ galaxies range from compact values of $\lesssim63$\,pc to large values of $\sim1.28$\,kpc, with a median and inner 68 percentile values of $R_e=310_{-160}^{+320}$\,pc. In Figure \ref{fig:size_hist}, we show the histogram of the size distribution in natural logarithmic scale following the literature.

We fit the galaxy size distribution with a log-normal function:
\begin{equation}\label{eq:lognorm}
    p(R_e){\rm d}R_e = \frac{1}{2\pi\sigma_{\ln R_e}} \exp\left(\frac{[\ln (R_e/\overline{R_e})]^2}{2\sigma_{\ln R_e}}\right) \frac{{\rm d}R_e}{R_e},
\end{equation}
where $\overline{R_e}$ and $\sigma_{\ln R_e}$ represent the $R_e$ value at the peak and standard deviation of the distribution, respectively. To account for the measurement uncertainties, we randomly draw 100 realizations based on the measured $R_e$ and the errors, repeating the fitting procedure. We obtain the best-fit median and inner 68 percentile $\overline{R_e}=0.245_{-0.004}^{+0.004}$\,kpc and $\sigma_{\ln R_e}=0.96_{-0.04}^{+0.04}$. These values are broadly consistent with the results from \citet{ono24,ono25}, who measured the galaxy sizes at $z=4-16$ in legacy fields with CEERS and JADES data. To test the robustness of our results, we repeat the analysis with $n$ fixed to 1.5 and find that our conclusions remain unchanged, with the smallest resolved effective radius being $R_e=73.6$\,pc. 

\subsection{Size - Luminosity relation}\label{subsec:size_lum}
\begin{figure}[ht!]
\begin{center}
\includegraphics[scale=0.6]{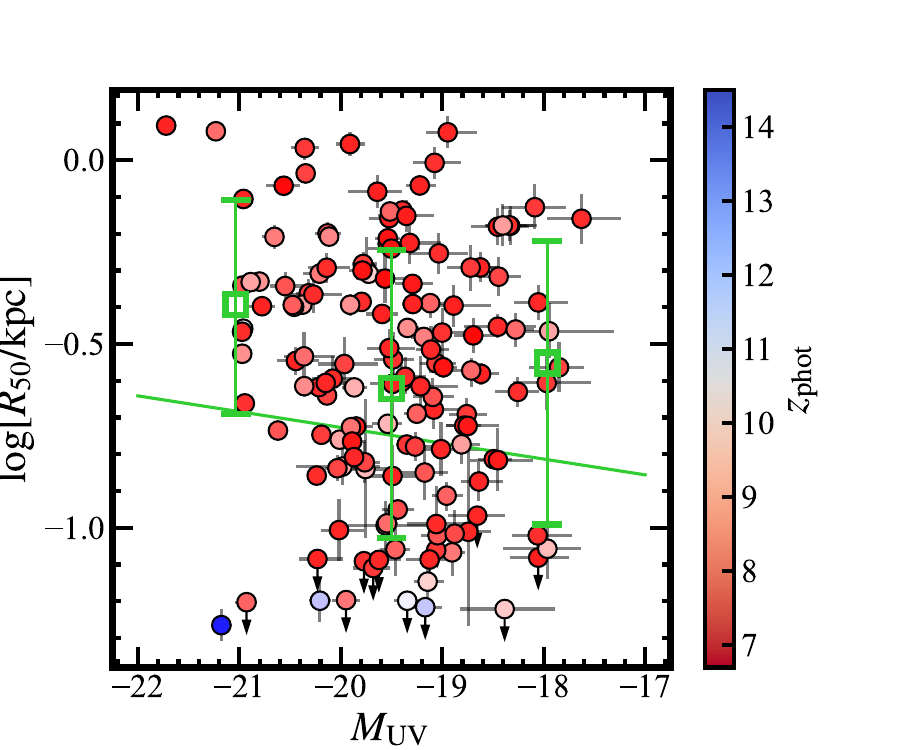}
\end{center}
\caption{Size-luminosity relation of our sample. All symbols and colors are the same as those in Figure \ref{fig:beta}.
}\label{fig:size}
\end{figure}

We investigate the relation between size and luminosity of $z=7-12$ galaxies. In Figure \ref{fig:size}, we present the $M_{\rm UV}-R_e$ relation of our sample, together with the median and inner 68 percentile $R_e$ in different $M_{\rm UV}$ bins. Following previous studies \citep[e.g.,][]{shibuya15,ono24}, we attempted to fit a power-law function to the observed $M_{\rm UV}-R_e$ relation:
\begin{equation}
    R_e = R_0\left(\frac{L_{\rm UV}}{L_0}\right)^\alpha,
\end{equation}
where $R_0$ is the effective radius at $L_0$ that corresponds to $-21$\,Mag. We conduct the fitting in logarithm space following the fitting procedures in Section \ref{subsec:uv}, obtaining the best-fit values of $R_0=0.34_{-0.04}^{+0.04}$\,kpc, $\alpha=-0.05_{-0.03}^{+0.03}$, and a large intrinsic scatter in logarithm of 0.31~dex. For $R_0$, our best-fit values are consistent with the results from \citet{ono25} at the similar redshift. However, we do not find correlation between $R_e$ and $L_{\rm UV}$, given that our best-fit $\alpha$ is close to zero and the intrinsic scatter is large. 

\begin{figure}[htb!]
\begin{center}
\includegraphics[scale=0.55]{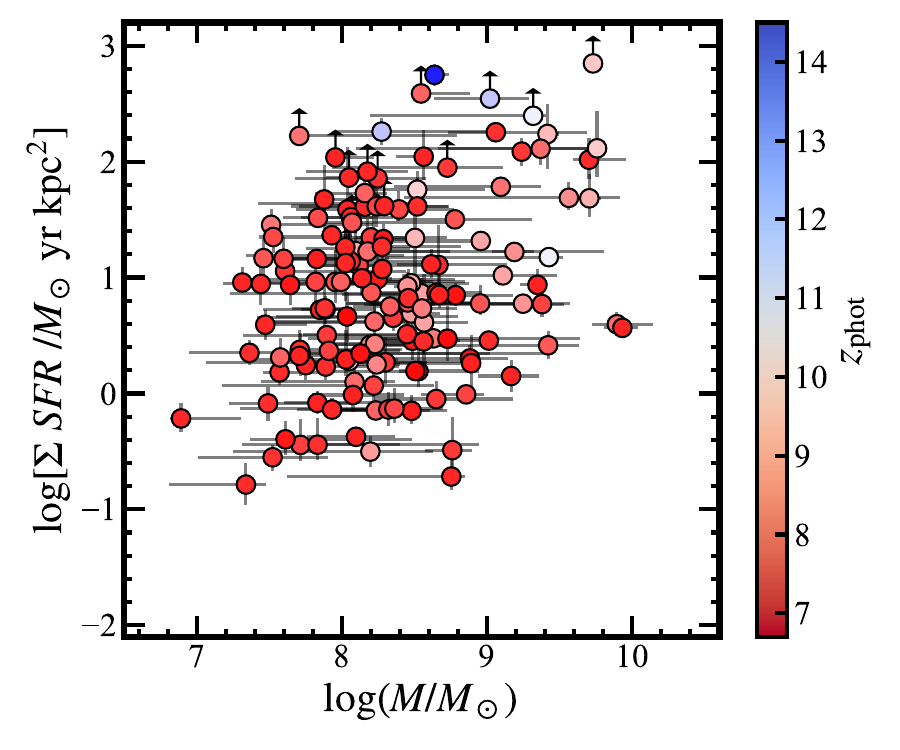}
\end{center}
\caption{The $\Sigma_{\rm SFR} - M_*$ relation of our sample. The symbols and colors are the same as those in Figure \ref{fig:beta}.
}\label{fig:ssfrd}
\end{figure}

Next, we derive the SFR surface density ($\Sigma_{\rm SFR}$) of our sample galaxies, which is estimated by:
\begin{equation}
    \Sigma_{\rm SFR} = \frac{0.5\,{\rm SFR}}{\pi R_e^2}.
\end{equation}
In Figure \ref{fig:ssfrd}, we show the relation between $M_*$ and $\Sigma_{\rm SFR}$
of our $z=7-14$ galaxy sample. Similar to previous JWST results \citep[e.g.,][]{calabro24,morishita24}, our sample features a wide range of $\Sigma_{\rm SFR}$ spanning $0.2 - 630\,M_\odot\,{\rm yr}^{-1}\,{\rm kpc}^{-2}$ with a median value of $\Sigma_{\rm SFR} = 6.6\,M_\odot\,{\rm yr}^{-1}\,{\rm kpc}^{-2}$, including $\sim 20\%$ with a large $\Sigma_{\rm SFR}$ of $\log [\Sigma_{\rm SFR}\,/\,{\rm M_\odot\,{\rm yr}^{-1}\,{\rm kpc}^{-2}}] > 1.5$. Interestingly, Figure \ref{fig:ssfrd} suggests that galaxies at higher redshift tend to have larger $\log\Sigma_{\rm SFR}$, which is consistent with the observed increase of $\log\Sigma_{\rm SFR}$ from $z=4-8$ \citep[e.g.,][]{calabro24}. The 20\% $M_{\rm UV}$ completeness limits for the F090W-, F115W-, and F150W-dropout galaxies are $-19.4$, $-19.8$, and $-20.0$, respectively.These modest differences in detection thresholds would produce only a $\sim0.2$~dex shift in $\Sigma_{\rm SFR}$. Hence, the observed trend of increased $\Sigma_{\rm SFR}$ towards higher redshifts is unlikely to be driven by selection effects. 
$\log\Sigma_{\rm SFR}$ has been shown to be related to the gas density of the interstellar medium \citep[ISM;][]{jiang19,reddy23}, ionization properties \citep{reddy22}, galactic feedback efficiency \citep[e.g.,][]{heckman16,llerena23}. 
The possible increase of $\log\Sigma_{\rm SFR}$ with redshift could result from denser environments, harder ionization, or varying outflow efficiency towards higher redshift. 

\section{Discussion} \label{sec:discuss}
\subsection{A Remarkably Bright, Compact Star-forming galaxy at $z=14$} \label{subsec:z14}
\begin{table}[ht!]
\centering
\caption{Physical Properties of beacon\_1420+5253\_4770} \label{tab:z14}
\begin{tabular}{ll}
\hline
\hline
\hline
$z_{\rm EAZY}$ & $13.52_{-0.14}^{+0.15}$ \\
$z_{\rm Prospect}$ & $13.71_{-0.15}^{+0.15}$ \\
$M_{\rm UV}$ & $-21.19_{-0.08}^{+0.08}$ \\
$\beta_{\rm UV}$ (photometry) & $-3.37_{-0.19}^{+0.24}$ \\
$\log(M_*/M_\odot)$ & $8.63_{-0.15}^{+0.31}$ \\
SFR (UV) & $10.51_{-0.68}^{+0.58}~M_\odot~{\rm yr}^{-1}$ \\
$R_e$ (UV) & $55.3_{-5.6}^{+5.6}$~pc \\
$\Sigma_{\rm SFR}$ (UV) & $547.4_{-116.1}^{+114.6}~M_\odot~{\rm yr}^{-1}~{\rm kpc}^{-2}$ \\
\hline
\end{tabular}
\end{table}

\begin{figure*}[ht!]
\begin{center}
\includegraphics[scale=0.58]{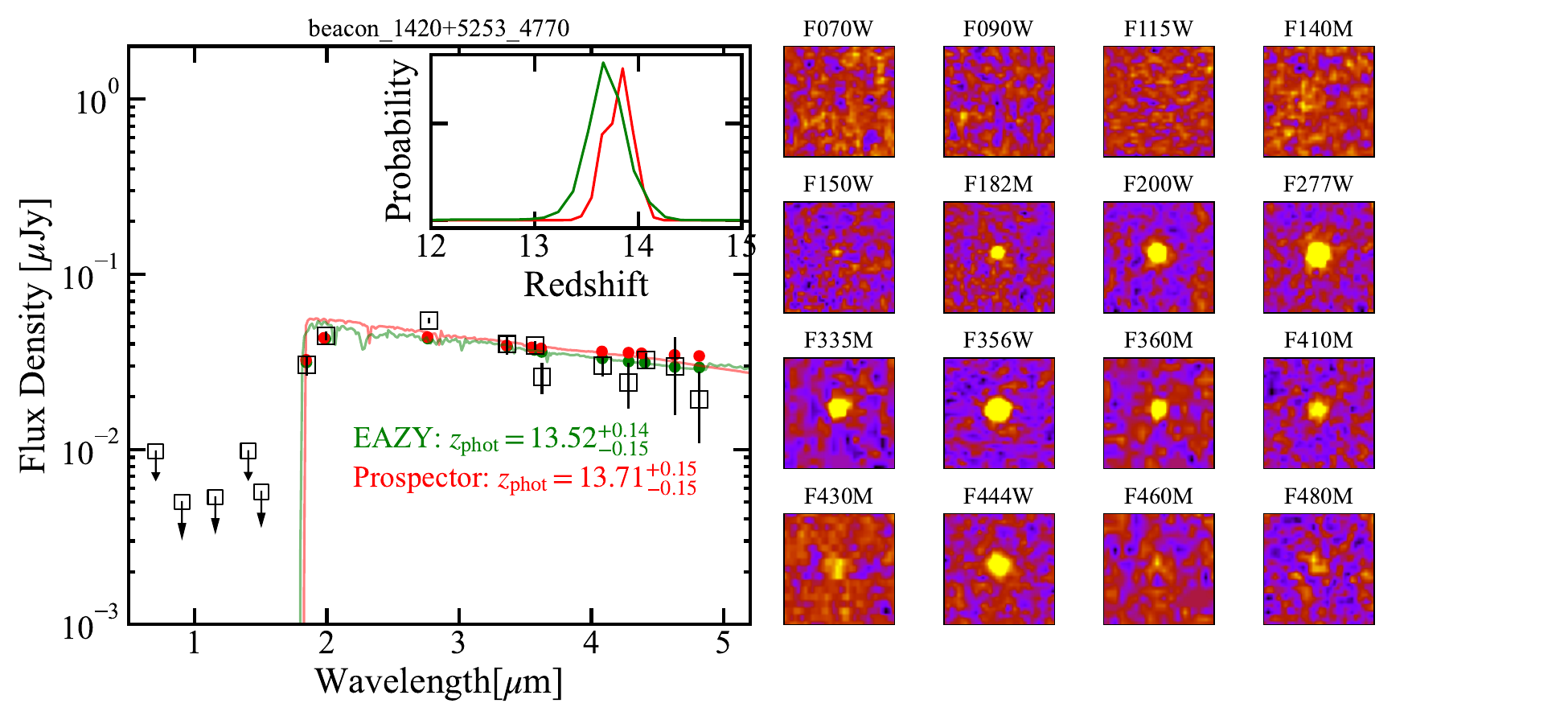}
\end{center}
\caption{Left: The best-fit \texttt{EAZY} (green) and \texttt{Prospector} (red) SEDs of beacon\_1420-5253\_4770, a candidate of brightest galaxy at $z>12$. The observed photometry is shown with black open squares with error bars. The photometry predicted from the best-fit SEDs is denoted with solid circles. The inset figure presents the $p(z)$ of the SED fitting results. Right: $1''\times1''$ image stamps of beacon\_1420-5253\_4770.
}\label{fig:z14}
\end{figure*}

In Section \ref{sec:sample}, we identified one F-150W dropout galaxy candidate, beacon\_1420+5253\_4770, at $z=13.7$ with an extremely bright UV luminosity of $M_{\rm UV} = -21.2$. As shown in Figure \ref{fig:z_muv}, if spectroscopically confirmed, it would be the brightest galaxy at $z>12$. Located in the EGS field, this object was also identified by \citet{weibel25} as ID89475. BEACON DR2 has one of the most filter coverages in this field with 16 broad and medium band filters, providing strong constraints for SED fitting. As presented in Figure \ref{fig:z14}, beacon\_1420-5253\_4770 is not detected in F140M and bluer filters. While there may be signals observed in the F150W filter, the detection is only at S/N$=1.5$. There is a clear break observed in the F182M filter with S/N$=7.8$, tightly constraining the photometric redshift of the object. 

The properties derived from BEACON DR2 photometry and SED fitting are summarized in Table \ref{tab:z14}. The galaxy exhibits a compact yet resolved morphology with an effective radius of $R_e=55.3\pm5.6$~pc, indicative of compact star formation with a high $\Sigma_{\rm SFR}$ of $547.4_{-116.1}^{+114.6}~M_\odot~{\rm yr}^{-1}~{\rm kpc}^{-2}$. However, the potential contribution from an AGN cannot be ruled out. The $\beta_{\rm UV}$ inferred from the photometry is remarkably blue, $-3.37_{-0.19}^{+0.24}$, likely induced by the high flux in the F277W filter. If the measured $\beta_{\rm UV}$ is true, beacon\_1420-5253\_4770 might have extremely low metallicity or high ionizing photon escape fraction with a weak nebular continuum \citep[e.g.,][]{yanagisawa24,topping24}. Alternatively, if the measured $\beta_{\rm UV}$ is affected by strong nebular emission lines (e.g., He{\sc ii}) in the F277W filter, this would imply the presence of powerful ionizing sources. Metal-poor stars, again, could explain such a scenario, while the existence of AGN remains another possibility. Future spectroscopic follow-up would help confirm the nature of beacon\_1420+5253\_4770.

\subsection{Galaxy Overdensities} \label{subsec:overdensity}
\begin{figure*}[ht!]
\begin{center}
\includegraphics[scale=0.58]{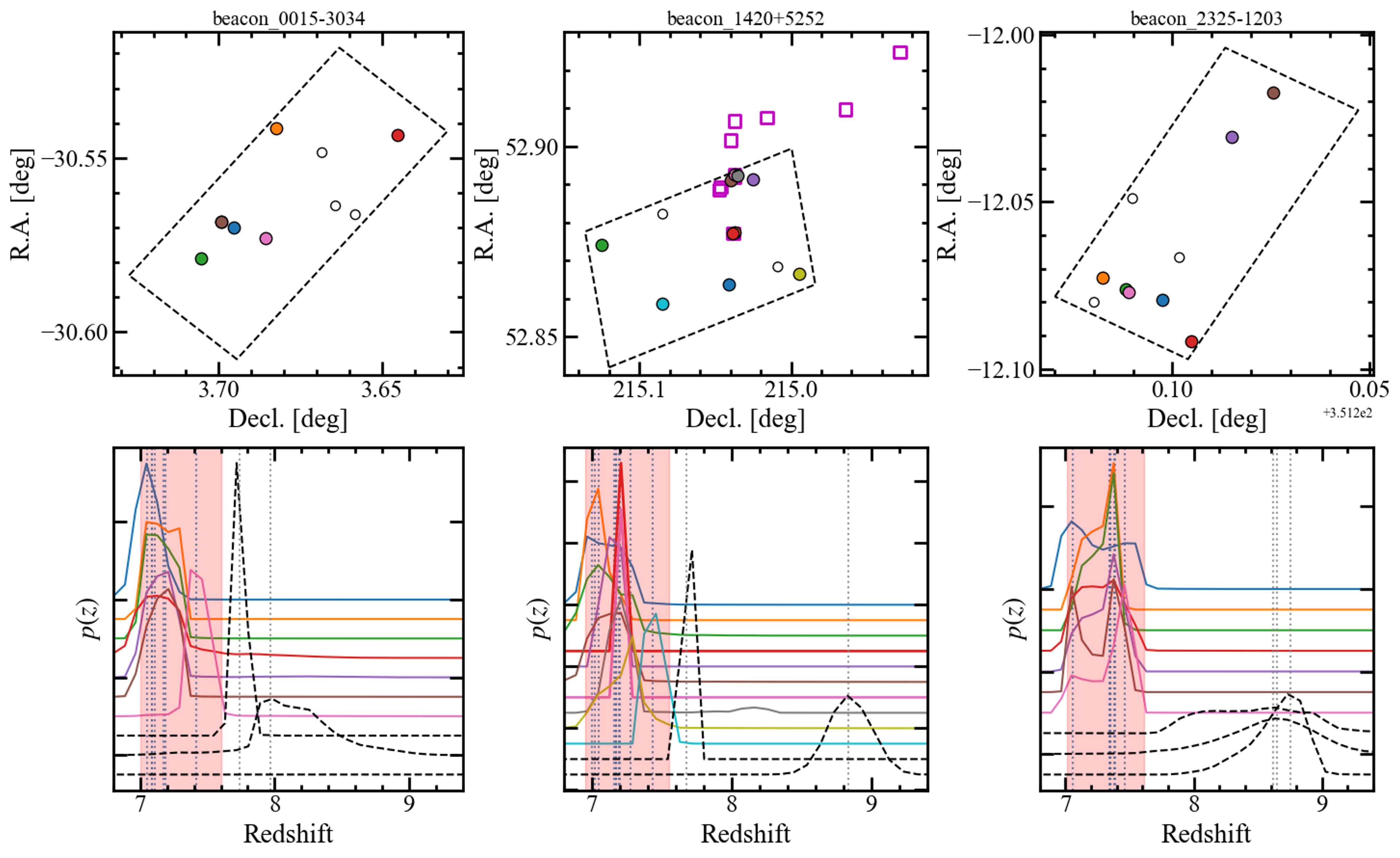}
\end{center}
\caption{The projected sky distribution (top) and $p(z)$ (bottom) of the three overdense fields identified in BEACON DR2. In the top row, the objects outside our $\delta z$ window are shown with open black points, while the NIRCam field of view (FoV) is indicated with black dashed boxes. For beacon\_1420+5252 in EGS field, the member galaxies of the $z=7.2$ overdensity in \citet{chen25} are indicated with magenta open squares. In the bottom row, the colors of $p(z)$ distributions match the colors of data points in the top row, except for galaxies outside the $\delta z$ window that are shown with black dashed curves. The red shaded regions indicate the $\delta z$ windows for identifying these overdensities. 
}\label{fig:od}
\end{figure*}
\begin{figure*}[htb!]
\begin{center}
\includegraphics[scale=0.58]{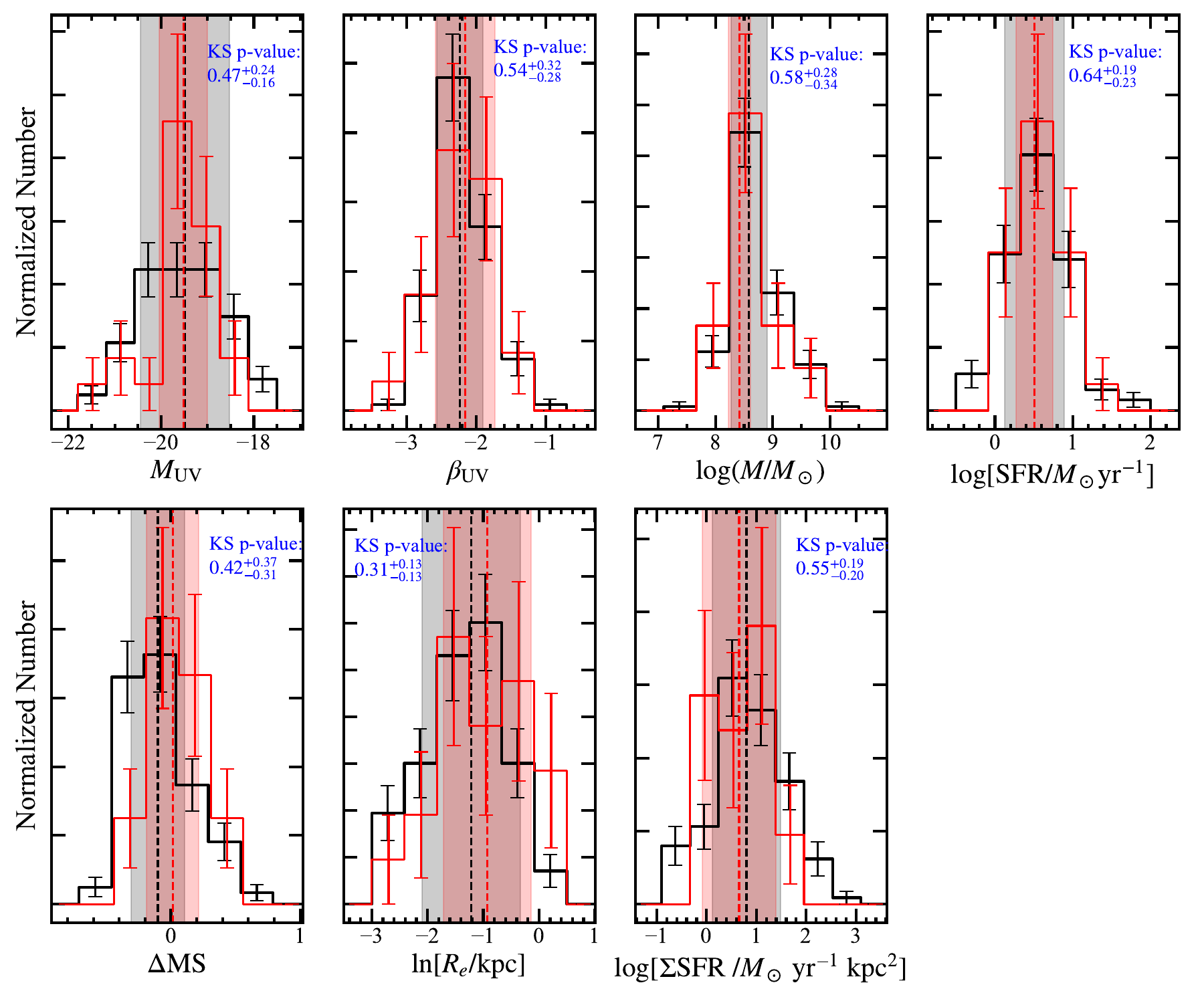}
\end{center}
\caption{Normalized histograms of $M_{\rm UV}$,$\beta_{\rm UV}$, $M_*$, SFR, sSFR, $\ln R_e$, and $\log\Sigma_{\rm SFR}$ of our galaxy sample in overdensities (red) and the field (black), with vertical bars indicating the Poisson errors. We perform the two-sample KS test on 100 Monte Carlo realizations of each physical parameter, and report the median and inner 68 percentile distribution of the p-values at the top of each panel (See Section \ref{subsec:overdensity}). 
}\label{fig:odhist}
\end{figure*}
One key advantage of wide-area, pure-parallel surveys such as BEACON is their ability to sample the universe through independent pointings, enabling robust identification of galaxy clustering. In this subsection, we identify and characterize the galaxy overdensities from the 36 BEACON DR2 pointings.

Following the methodology of \citet{trenti12}, as detailed in K. C. Kreilgaard et al. (in preparation), we evaluate the statistical significance of an overdense field by computing the Poisson probability of observing $N_{\rm obs}$ or more galaxies in a pointing within a certain redshift window where $N_{\rm exp}$ are expected, $P(N\geq N_{\rm obs}\,|\,N_{\rm exp})$. Within each search volume, we obtain $N_{\rm obs}$ by counting the galaxies whose best-fit \texttt{EAZY} photometric redshifts, evaluated at the peak of $p(z)$, fall within the redshift window. The \texttt{EAZY} photometric redshift is chosen over the \texttt{Prospector} results to maintain consistency with the simulations used to derive $N_{\rm exp}$. 
We calculate $N_{\rm exp}$ by integrating the UV luminosity functions (LFs) derived from the same BEACON DR2 dataset across 36 pointings (K. C. Kreilgaard et al., in preparation) over the relevant redshift interval. The integration limit in $M_{\rm UV}$ is set by the faintest member galaxy identified in each search volume, which varies among different pointings and redshift windows and may fall below the 100\% detection completeness. However, such an effect is incorporated in the $N_{\rm exp}$ estimation through the simulated completeness function and thus the UV LF determination. Given the 100\% purity of our sample selection (Section \ref{sec:sample}), we do not consider contamination in our $N_{\rm exp}$ estimation.

We search for overdensities with the redshift window of $\delta z=0.6$. This large range is chosen to match the accuracy of the \texttt{EAZY} photometric redshift in Section \ref{sec:sample} ($\sim2\sigma_{\rm MAE}$). We identify a field as an overdensity candidate if $P(N\geq N_{\rm obs}\,|\,N_{\rm exp})<0.003$, corresponding to a $3\sigma$ detection. To further account for the uncertainties in \texttt{EAZY} photometric redshifts, we also calculate the probability that all the member galaxies fall within $\delta z$ window, $P_{\rm OD}$, based on their $p(z)$ distributions. We require $P_{\rm OD}>0.5$ to classify an overdensity as a robust detection. With the $P(N\geq N_{\rm obs}\,|\,N_{\rm exp})<0.003$ and $P_{\rm OD}>0.5$ criteria, we identify three galaxy overdensity fields, listed in Table \ref{tab:od}. In Figure \ref{fig:od}, we show the sky distribution and the \texttt{EAZY} $p(z)$ of these fields. Below we briefly describe these three fields:

\textit{beacon\_0015-3034} This overdense field is close to the ABELL-2744 field, where a $z=7.88$ protocluster has been spectroscopically confirmed \citep[e.g.,][]{morishita25od}. However, no spectroscopically confirmed galaxies are found in our BEACON pointing. The field also lacks bright galaxies with $M_{\rm UV}<20.5$. Two galaxies exhibit relatively red UV continua with $\beta_{\rm UV}>-1.7$.

\textit{beacon\_1420+5252} Located within EGS field, this overdense region partially overlaps with a $z=7.3$ protocluster spectroscopically identified by \citet{chen25}, with 2/10 of our member galaxies included in their sample. The field contains two bright galaxies with $M_{\rm UV}<20.5$, which may be associated with the large ionizing bubbles identified in the EGS field \citep[e.g.,][]{tang23,chen25}. One of the two bright galaxies has a relatively large stellar mass of $\log(M_*/M_\odot)>9.5$. There are another two red galaxies with $\beta_{\rm UV}>-1.7$, suggesting a comparatively evolved environment in this overdense field. 

\textit{beacon\_2325-1203} This field is in a previously unexplored extragalactic region, with BEACON providing the first photometric data coverage. The field contains one bright galaxy with $M_{\rm UV}<20.5$, but no galaxies having red UV continua or large stellar masses ($\log(M_*/M_\odot)>9.5$).



\begin{deluxetable}{cccc}
\tablewidth{0pt}
\tablecaption{Summary of the three galaxy overdensity fields found in BEACON DR2 \label{tab:od}}
\tablehead{
\colhead{Field} & \colhead{$n$} & \colhead{$\delta$} & \colhead{$z_{\rm med}$}
}
\startdata
beacon\_0015-3034 & 7 & 13.4 & 7.18 \\
beacon\_1420+5252 & 10 & 26.9 & 7.17 \\
beacon\_2325-1203 & 7 & 12.9 & 7.38 \\
\enddata
\end{deluxetable}

To explore how galaxy properties vary with environment at $z>7$, we divide our galaxy sample into two subsets, the 24 galaxies in overdense fields listed in Table \ref{tab:od} (hereafter ``overdense galaxies'') and those outside these regions (hereafter ``field galaxies"). To make a fair comparison, we limit the field galaxy sample to F090-dropout galaxies. The total number of field galaxies is $151-24=127$.
In Figure \ref{fig:odhist}, we present the normalized histograms of physical properties, including $M_{\rm UV}$,$\beta_{\rm UV}$, $M_*$, SFR, galaxy distance to the main sequence ($\Delta$MS, Figure \ref{fig:ms}), $\ln R_e$, and $\log\Sigma_{\rm SFR}$, for overdense and field galaxies, with Poisson-based error bars. Histogram bin widths are chosen to exceed typical measurement uncertainties. The median and inner 68th percentile of each distribution are indicated by vertical dashed lines and shaded regions, respectively. Visually, the $\Delta$MS distribution for overdense galaxies appears slightly skewed toward higher values compared with field galaxies, although the medians of the two distributions remain consistent within statistical uncertainties. 
For all other properties, the distributions of overdense and field galaxies appear consistent. 

To quantitatively assess these differences, we perform two-sample Kolmogorov–Smirnov (KS) tests, incorporating measurement uncertainties through 500 Monte Carlo realizations, and report the median and inner 68th percentile of the resulting $p$-value distributions in Figure \ref{fig:odhist}. We find that the median $p$-values for all examined properties are well above 0.05 (corresponding to a 2$\sigma$ significance threshold), indicating that none of the parameters show statistically significant differences between overdense and field galaxies.

In a previous JWST study, \citet{li25od} investigated the dependence of galaxies properties on environment based on 26 overdensities at $z<7$, finding that overdense galaxies may experience accelerated star formation and dust enrichment as indicated by higher specific SFR (sSFR) and redder $\beta_{\rm UV}$, respectively. In contrast, we do not find such trends in Figure \ref{fig:odhist}. This difference might arises because our analyses focus on a slightly higher redshift regime at $z>7$, where the Universe is yonger than 800~Myr. Although elevated sSFRs and/or redder $\beta_{\rm UV}$ have been reported in individual extreme overdensities at at $z>7$ \citep[e.g.,][]{daikuhara25,witten25}, our results based on 35 independent BEACON DR2 pointings suggest such environmental effects may not be widespread until later cosmic epochs.

Another plausible explanation for the lack of a detected dependence of sSFR (or $\Delta{\rm MS}$) on environment is the limited sample size. To assess the statistical power of our data, we perform simulations to determine the sample size required to detect a difference in the $\Delta{\rm MS}$ distributions between overdense and field galaxies. Assuming that the underlying distributions of $M_*$ and SFR is identical to the current sample, we generate bootstrap realizations in which the numbers of overdense and field galaxies are increased by a factor of $N$, and repeat the same analysis. We find that with $10\times$ more galaxies, the difference in $\Delta{\rm MS}$ between overdense and field galaxies would be recovered with a median $p$-value of 0.017 ($\sim2.4\sigma$), highlighting the need for future pure-parallel observations.
Future spectroscopic follow-up observations would also help improve the measurement uncertainties and provide constraints on the environmental effects of various galaxy physical parameters.

\section{Summary} \label{sec:summary}
In this work, we present the analysis of 161 robust galaxy candidates at $z=7-14$ selected from 36 independent pointings (corresponding to $\sim350$\,arcmin$^2$) in the second data release (DR2) of BEACON, a JWST Cycle 2 pure-parallel NIRCam imaging program. We estimate the properties, including $z_{\rm phot}$ and $M_*$, of our sample based on SED fitting using \texttt{Prospector}, utilizing the redshift probability distribution obtained from \texttt{EAZY}. Our sample contains 14 spectroscopically confirmed sources, among which the $z_{\rm phot}$ errors, defined as $|z_{\rm phot}-z_{\rm spec}|/(1+z_{\rm spec})$, are smaller than $10\%$. Out of the 161 objects, 146 are selected as F090W-dropout galaxies ($7.3\lesssim z \lesssim 9.7$), 14 as F115W-dropout galaxies ($9.7\lesssim z \lesssim 13$), and one as a F150W- dropout galaxy ($13\lesssim z \lesssim 18$).

Combining our SED analysis with multi-band photometry, we characterize the key physical properties of our sample, including $M_{\rm UV}$, $\beta_{\rm UV}$, $M_*$, and SFR. The UV luminosity functions and clustering properties will be presented in the accompanying paper (Kreilgaard et al. in preparation). We find no correlation between $\beta_{\rm UV}$ and $M_{\rm UV}$, whose linear correlation has a slope of $\frac{{\rm d}\beta_{\rm UV}}{{\rm d}M_{\rm UV}} = -0.01_{-0.04}^{+0.04}$. In contrast, the $\beta_{\rm UV}-M_*$ relation has a linear slope of $\frac{{\rm d}\beta_{\rm UV}}{{\rm d}M_*} = 0.27_{-0.04}^{+0.04}$. The $M_*-$SFR main sequence of our sample is consistent with previous measurements at the similar redshift \citep{clarke25,morishita24,rb24}, suggesting no differences between the results from pure-parallal and legacy fields.

We analyzed the rest-frame UV morphology of our galaxies with NIRCam imaging data. Assuming $n=1$ S\^ersic profiles, the sizes of our galaxies span $R_e=0.063-1.28$\,kpc, and can be described by a log-normal distribution that peaks at $\overline{R_e}=0.245_{-0.004}^{+0.004}$\,kpc with a standard distribution of $\sigma_{\ln R_e}=0.96_{-0.04}^{+0.04}$.

We highlight the single F150W-dropout galaxy in our sample, beacon\_1420+5253\_4770, at $z=13.7$ with $M_{\rm UV}=-21.2$ that is brighter than any spectroscopically confirmed galaxies at $z>12$. This galaxy has a compact but resolved morphology with $R_e=55.3\pm5.6$~pc, a remarkably high $\Sigma_{\rm SFR}$ of $547.4_{-116.3}^{+114.6}~M_\odot~{\rm yr}^{-1}~{\rm kpc}^{-2}$, and an extremely blue UV continuum of $\beta_{\rm UV}=-3.37_{-0.19}^{+0.24}$ measured from photometry, suggesting that this object might have an extremely low metallicity, a high ionizing gas escape fraction, or contribution from AGN.

Utilizing the multiple independent sightlines of BEACON DR2, we identified three galaxy overdensity fields at $z=7.0-7.5$ with $>3\sigma$ significance levels. One of these fields, beacon\_1420+5252, partially overlaps with a spectroscopically confirmed overdensity at $z=7.2$ in EGS field \citep{chen25}. The other two fields are newly identified in this study, showcasing the strength of pure-parallel imaging in identifying clustering of galaxies. We further investigate the dependence of different physical parameters on various environment, dividing our galaxies into those located in (``overdense galaxies'') and out of (``field galaxies'') the overdense fields. We find that there is no significant differences in the distributions of these parameters, suggesting that the accelerated star formation  found in individual systems \citep[e.g.,][]{witten25,daikuhara25} may not be widespread until later epochs \citep{li25od}. Alternatively, the absence of a detectable environmental dependence may simply reflect the limited sample size. Our simulations indicate that an enhancement of SFR in overdense galaxies would be detected at a significance level exceeding $2\sigma$ if the current sample were increased by an order of magnitude. Future pure-parallel observations will therefore be crucial for probing the environmental dependence of galaxy properties at cosmic dawn.

\begin{acknowledgments}
Support for program 3990 was provided by NASA through the Space Telescope Science Institute, which is operated by the Association of Universities for Research in Astronomy, Inc., under NASA contract NAS 5-03127. 
All of the data presented in this paper were obtained from the Mikulski Archive for Space Telescopes (MAST) at the Space Telescope Science Institute. 
Some of the data products presented herein were retrieved from the Dawn JWST Archive (DJA). DJA is an initiative of the Cosmic Dawn Center (DAWN), which is funded by the Danish National Research Foundation under grant DNRF140. MB acknowledges support from the ERC Grant FIRSTLIGHT and Slovenian national research agency ARIS through grants N1-0238 and P1-0188.

\end{acknowledgments}




\facilities{JWST(NIRCam)}

\software{astropy \citep{2013A&A...558A..33A,2018AJ....156..123A,2022ApJ...935..167A}, EAZY \citep{brammer08}, Galfit \citep{peng02,peng10}, Prospector \citep{prospect}, 
          Source Extractor \citep{1996A&AS..117..393B}
          }



\bibliography{sample7}{}
\bibliographystyle{aasjournalv7}

\end{document}